\documentclass[aps,prl,superscriptaddress,groupedaddress,floatfix,longbibliography]{revtex4-1}
\usepackage[utf8]{inputenc}

\usepackage{graphicx}  
\usepackage{dcolumn}   
\usepackage{bm}        
\usepackage{amssymb}   
\usepackage{amsmath}   
\usepackage{amsthm}    
\usepackage{xcolor}    
\usepackage{tikz}      
\usepackage{ifthen}    
\usepackage{subfigure}
\usepackage{xspace}
\usepackage{lipsum}
\usepackage{gensymb}
\usepackage{appendix}
\usepackage{hyperref}
\usepackage{soul}
\usepackage{lineno}

\usepackage[normalem]{ulem}
\newcommand{\xtec}{\textit{X-TEC}}
\newcommand{\pdx}{Pd$_x$ErTe$_{3}$}

\begin{document}
\title{Bragg glass signatures in PdxErTe3 with X-ray diffraction Temperature Clustering (X-TEC)} 

\author{Krishnanand Mallayya$^1$,  Joshua Straquadine$^2$,  Matthew J. Krogstad$^{3,4}$, Maja D. Bachmann$^2$, Anisha G. Singh$^2$, Raymond Osborn$^3$, Stephan Rosenkranz$^3$, Ian R. Fisher$^2$, Eun-Ah Kim$^{1,5 \ast}$\\
\it{$^{1}$Department of Physics, Cornell University, Ithaca, NY 14853, USA}\\
\it{$^{2}$Geballe Laboratory for Advanced Materials and Department of Applied Physics,
Stanford University, Stanford, California 94305, USA\\
and Stanford Institute for Materials and Energy Sciences, SLAC National Accelerator Laboratory,
2575 Sand Hill Road, Menlo Park, California 94025, USA}\\
\it{$^{3}$Materials Science Division, Argonne National Laboratory, Lemont, IL 60439, USA}\\
\it{$^{4}$Advanced Photon Source, Argonne National Laboratory, Lemont, IL 60439, USA}\\
\it{$^{5}$Department of Physics, Ewha Womans University, Seoul 03760, South Korea}\\
\it{$^\ast$To whom correspondence should be addressed; E-mail:  eun-ah.kim@cornell.edu.}
}

\date{\today}

\baselineskip24pt

\begin{abstract}
The Bragg glass phase is a nearly perfect crystal with glassy features predicted to occur in vortex lattices and charge density wave systems in the presence of disorder. Detecting it has been challenging despite its sharp theoretical definition in terms of diverging correlation lengths. Here, we present evidence supporting a Bragg glass phase in the systematically disordered charge density wave material PdxErTe3. We do this using comprehensive x-ray data and a machine learning analysis tool called X-ray temperature clustering, or X-TEC. We establish a diverging correlation length in samples with moderate intercalation over a wide temperature range. To enable this analysis, we introduced a high-throughput measure of inverse correlation length that we call peak spread. The detection of Bragg glass order and the resulting phase diagram advance our understanding of the complex interplay between disorder and fluctuations significantly. Moreover, the use of our analysis technique to target fluctuations through a high-throughput measure of peak spread can revolutionize the study of fluctuations in scattering experiments.

\end{abstract}
\maketitle
\section{Introduction}

The interplay between disorder and fluctuations can result in complex new phases, such as spin glasses, recently celebrated through a Nobel prize~\cite{NobelPrize.org}. While theoretical frameworks for understanding such complex phases can have far reaching implications, it can be challenging to untangle the subtleties of this interplay from experimental data when finite experimental resolution and noise obscure comparisons with idealized theoretical predictions. The Bragg glass is an example of such an elusive novel phase~\cite{Giamarchi1994PRL,Giamarchi1995Phys.Rev.B}. It is an algebraically ordered glass, which can appear as ordered as a perfect crystal, but whose Bragg peak intensities diverge as power-laws~\cite{Nattermann1990Scaling,Giamarchi1994PRL,Gingras1996Phys.Rev.B,Kierfeld1997Phys.Rev.B,Fisher1997Phys.Rev.Lett.,Giamarchi1997Phys.Rev.B,Rosso2003Phys.Rev.B,Rosso2004Phys.Rev.B,Brazovskii2004Adv.Phys.,doussal2011novel,Mross2015Phys.Rev.Xa}. However, with the experimental resolution cutting off the divergence in actual data, it can be challenging to detect a Bragg glass. While the Bragg glass has been proposed for charge-density-wave systems (CDW)~\cite{Rosso2003Phys.Rev.B,Rosso2004Phys.Rev.B} and vortex lattices~\cite{Bogner2001Phys.Rev.B}, unambiguous direct evidence has so far been limited to vortex lattices~\cite{Klein2001Nature,Bogner2001,Park2003Phys.Rev.Lett.,Daniilidis2007Phys.Rev.Lett.a,Toft_Petersen2018NatCommuna}. Since the lattice periodicity in a vortex lattice is controlled by magnetic field, \textcite{Klein2001Nature} could rely on the magnetic field-independent width of the rocking curve as evidence of the absence of an intrinsic length scale and the underlying algebraic order. An STM probe on 1T-TaS$_2$ also revealed the  Bragg glass decay of translational order in a vortex lattice of CDW topological defects~\cite{Altvater2021Appl.Phys.Lett.}. However, evidence of Bragg glass phenomena in incommensurate CDW systems with emergent, system-specific periodicity is suggestive at best and
limited to scanning tunneling microscopy (STM) studies of NbSe$_{2}$~\cite{Okamoto2015Phys.Rev.Lett.} and  \pdx~\cite{Fang2019Phys.Rev.B}. Hence, whether an algebraically ordered CDW phase can exist as a bulk phase in a CDW system or whether CDW's always respond to disorder as, for example, a vestigial nematic with a short correlation length~\cite{Nie2014PNAS} remains an open problem.

In this work, we present the first bulk signature supporting a Bragg glass phase in a systematically disordered CDW material, \pdx, using comprehensive single-crystal x-ray scattering and a novel ML-based method of data analysis called X-ray Temperature Clustering (\xtec) [introduced by some of us in Ref~\citenum{Venderley2022Proc.Natl.Acad.Sci.}]. Specifically, we provide evidence indicating the vanishing intrinsic length scale by 
tracking the temperature and momentum dependence of all the CDW peaks in a reciprocal space volume spanning ~20000 Brillouin Zones (BZ) with the help of \xtec. 
To the best of our knowledge, this is the first time CDW fluctuations have been analyzed from more than a handful of peaks. The statistics afforded by such an unprecedented comprehensive analysis of the CDW peak width
enable an accurate assessment of CDW correlation lengths by eliminating contributions to the observed peak width from crystal imperfections, and statistically minimizing errors near the resolution limit.  The resulting phase diagram establishes the Bragg glass to be the dominant phase, aligning with the onset of the transport anisotropy previously observed~\cite{Straquadine2019Phys.Rev.B}.

The notion that a quasi-long range ordering of vortex lattices and charge density waves is possible in the form of Bragg glass in the presence of disorder~\cite{Nattermann1990Scaling,Giamarchi1994PRL,Giamarchi1995Phys.Rev.B}  was a surprising theoretical prediction contradicting the long-standing wisdom that order parameters that couple linearly with a disorder potential are destined to be short-ranged at best~\cite{Imry1975Phys.Rev.Lett.}. The key difference lies in whether the phase fluctuation is allowed to grow indefinitely or the phase is defined compactly within $[0,2\pi)$. When considering a phase linearly coupling to the disorder potential, Imry and Ma showed that phase fluctuations can grow arbitrarily to overcome the elastic restoring energy, resulting in short-range correlations in dimensions below 4D~\cite{Imry1975Phys.Rev.Lett.}. 
However, \textcite{Nattermann1990Scaling} noted that the phase of periodic states such as charge density waves and vortex lattices should be defined compactly within $[0,2\pi)$; this compactness keeps the impact of the disorder potential in check. Specifically,  the disorder-averaged potential energy depends on the exponential of the phase fluctuations, allowing for quasi-long range order in the phase correlations in 3D~\cite{Nattermann1990Scaling,Giamarchi1994PRL,Giamarchi1995Phys.Rev.B} (section A in SM). Evidence for the divergence of the correlation length with such quasi-long range order would be the vanishing width of structure factor peaks associated with the periodicity\cite{Rosso2003Phys.Rev.B,Rosso2004Phys.Rev.B}.
Since the vortex or CDW phase is well-defined within $[0,2\pi)$ only in the absence of dislocations~\cite{Gingras1996Phys.Rev.B,Kierfeld1997Phys.Rev.B,Fisher1997Phys.Rev.Lett.}, observation of such an absence\cite{Okamoto2015Phys.Rev.Lett.,Fang2019Phys.Rev.B} establishes a necessary, but not a sufficient, condition for a Bragg glass.

There are many challenges in making direct observations of Bragg glass phenomena in CDW systems. First, for a systematic understanding of the role of disorder, a material family with adjustable disorder is needed. 
Second, for a statistically significant separation of real-life issues such as
crystal imperfections and finite resolution 
all contributing to the peak width from the sought-after fluctuation effects, 
a large volume of comprehensive data is a must.
Finally, for a reliable analysis of such large volumes of comprehensive data, a new approach to the data analysis is critical.  
We turn to \pdx\ to address the first, material system challenge~\cite{Straquadine2019Phys.Rev.B,Fang2019Phys.Rev.B,Fang2020Phys.Rev.Research}. Pristine ErTe$_3$ is a member of the rare earth tritelluride family with nearly square Te nets and a glide plane distinguishing the two in-plane directions $a$ and $c$ [Fig.~1(a)]. It hosts a unidirectional CDW ordering  (CDW-1, along $c$ axis) below a critical temperature $T_{c1}$ and an orthogonal unidirectional CDW (CDW-2, along $a$ axis) below $T_{c2}$, where  $T_{c2}<T_{c1}$ due to a weak orthorhombicity~\cite{Ru2008Phys.Rev.B}. Pd intercalation provides localized disorder potentials at random sites, making \pdx\ a model system to study emergent phases from  suppression of long-range CDW order~\cite{Straquadine2019Phys.Rev.B,Fang2019Phys.Rev.B}. Transport measurements in the pristine sample  have revealed the onset of anisotropy in resistivity between the $a$ and $c$ axis at the CDW transition temperature~\cite{Sinchenko2014PRL,Straquadine2019Phys.Rev.B}. Increasing intercalation lowers the onset temperature for the transport anisotropy~\cite{Straquadine2019Phys.Rev.B}. While this reveals the broken $C_4$ symmetry, two possible candidates for the disordered CDW phase remain open: a short range ordered CDW forming a vestigial nematic phase pinned by weak symmetry-breaking field and a Bragg glass phase characterized by quasi-long range CDW order with divergent power-law correlations.

We overcome the second challenge of the data volume by taking  x-ray temperature series data for single crystals  \pdx\ at different intercalation strengths ($x=0.0,~0.5\%,~2.0\%,~2.6\%, ~2.9\%$). We utilize highly efficient methods for collecting total x-ray scattering over large volumes of reciprocal space recently developed on Sector 6-ID-D at the Advanced Photon Source \cite{Krogstad2019tc}. In each measurement, a crystal is rotated continuously through 360\degree at a rate of 1\degree/s while images are collected on a fast area detector (Pilatus 2M CdTe) every 0.1~s,
with a monochromatic incident x-ray energy of 87~keV. Three rotations are required to fill in gaps between the detector chips. Uncompressed, the raw data volume is over 100~GB. 
While the data volume is reduced by 
an order-of-magnitude after transforming the images into reciprocal space meshes,
these meshes  include over 10,000 Brillouin Zones (BZ) and approximately $5 \times 10^8$ bins containing data. Such volumes are collected at a series of temperatures from 30~K to 300~K, controlled by a helium/nitrogen cryostream. During the transformation  of the images to the reciprocal space, one should ensure that the orientation is properly maintained at each temperature (SM section K).

Finally, we overcome the challenge of data analysis through a scalable extraction of theoretically relevant features using a machine learning algorithm~\xtec~\cite{Venderley2022Proc.Natl.Acad.Sci.}.  
In the x-ray scattering data, the CDW lattice distortions are manifest as satellite peaks around each of the Bragg peaks [Fig.~1(b)].
We focus on the temperature evolution of two features associated with the CDW peaks [illustrated in Fig.1 (c)]: the peak height and the peak width. In the long-range ordered CDW of the pristine sample, the temperature dependence of peak heights are sufficient to reveal the 
 order parameter and the transition temperature $(T_c)$. However, disorder can often broaden the transition. Moreover, in Bragg glass, the temperature dependence of the peak height does not reveal a clear onset behavior. This is because even after the breakdown of Bragg glass order, a non-vanishing superlattice peak intensity continues to persist to higher temperatures due to thermal and disorder fluctuations to CDW order [see the first row of the table in Fig.~1(d)].
On the other hand, the peak width of the CDW peaks [$\Gamma$ in Fig.~1(c)] should vanish upon transition into both the 
 long-range ordered and Bragg glass phases [see the second row of the table in Fig.~1(d)], whereas a short-range ordered phase, such as a vestigial nematic, will show a finite width down to the lowest temperatures (section C in SM). Invariably, finite experimental resolution and finite amount of crystalline defects present in samples will mask this difference. 
 However, with enough statistics over a range of temperatures, the temperature dependence of the width can be extrapolated to the vanishing point and allow for the determination of the Bragg glass transition temperature $T_{BG}$ 
[see the second row in Fig.~1(d)]. Hence, the Bragg glass phase is a phase that appears like long-range ordered phase despite presence of weak disorder. The effect of disorder can be evident from the asymmetry between a pair of CDW satellite peaks. As shown in the section D in SM, disorder pinning is required for such asymmetry~\cite{Ravy1993J.Phys.IVFrance}. Therefore we anticipate asymmetric pairs of CDW peaks with vanishing width in a Bragg glass phase. 

Manually tracking the three features of the disordered CDW from large data sets [Fig.~1(e)] presents a daunting challenge, hence the need for an automated machine learning approach like \xtec. At the core of the \xtec\ algorithm is a Gaussian Mixture Model (GMM) clustering to identify distinct temperature trajectories from the x-ray data. This is achieved by representing the intensity-temperature trajectory at each $\vec{q}$ in reciprocal space: $\{I_{\vec{q}}(T_i)\}$ spanning $d$ number of temperatures $\{T_1, \dots,T_d\}$  [Fig.~1(e) with $d=19$], as a point in the $d$-dimensional space [see  SM-Fig.~(2) in SM section F, for a 2D projection of this space]. From this distribution in hyperspace, GMM identifies a number of distinct clusters and assigns points to each one. From these cluster assignments, we can identify distinct intensity-temperature trajectories present in the data [Fig.~1(g)], thereby revealing the physically interesting ones, such as those representing the temperature dependence of the order parameters.

We first benchmark the \xtec\ outcomes against known results for the pristine ErTe$_3$ data [Fig.~2(a)]. The  collection of raw data fed into \xtec\ yields two well-defined CDW transitions in a matter of minutes [see section F of SM for details on \xtec\ processing]. From the mean  trajectories of the intensities in these two clusters, we can identify two transition temperatures $T_{c1}\approx 260$K and $T_{c_2}\approx 135$K. The transition temperatures identified by \xtec\ are consistent with those determined from transport anisotropies~\cite{Straquadine2019Phys.Rev.B}. Turning to where the clusters are located in reciprocal space, we find that the two intensity clusters correspond to the CDW-1 and CDW-2 peaks, whose  $K$-dependence are consistent with known selection rules [Fig.~2(b-c)].
Both the CDW-1 and CDW-2 peaks are sharp, as expected for 3D CDW order, and therefore satisfy the dimensionality condition necessary for a stable Bragg glass phase~\cite{Zeng1999Phys.Rev.Lett.,LeDoussal2000PhysicaC:Superconductivity}. In the rest of the paper, we focus on the CDW-1 peaks with higher transition temperature matching the expectations of the BCS order parameter in the pristine sample [Fig.~2(d)].  

Repeating the \xtec\ analysis on all intercalated samples, one readily extracts our first feature of interest: the temperature trajectory of the CDW peak intensity (peak height) shown in Fig.~2(d).  We show the average trajectory of all CDW-1 peaks at various intercalation levels.  Increasing intercalation suppress the overall intensity of modulations but more importantly it spreads out the intensity distribution as a function of temperature, leaving a long tail up to higher temperatures.  The long tail due to pinned CDW fluctuations at the intercalants ~\cite{Rouziere2000Friedel} hinders a clean determination of the putative Bragg glass  transition ~\cite{Nattermann1990Scaling,Giamarchi1994PRL,Giamarchi1995Phys.Rev.B} from the temperature dependence of the peak intensities. On the other hand, an STM study on \pdx~\cite{Fang2019Phys.Rev.B} has shown the absence of free dislocations, which is necessary for a Bragg glass, for moderate levels of intercalation $(x\lesssim 2\%)$ at a base temperature $\lesssim 1.7$ K.

To target the unambiguous signature of a Bragg glass, we now turn to our second feature of interest, namely the peak width. The objective is to separate three sources of CDW peak broadening with confidence: (1) the instrumental resolution, (2) finite CDW correlation lengths, (3) crystal imperfections.
Our approach is to use the ${\vec{q}\equiv(H,K,L)}$ dependence  of the peak broadening since only crystal imperfections would
result in a (quadratic) momentum dependence across BZ's (section E in SM).  This strategy requires measuring peak widths over a statistically significant number of BZ's.  While our experimental setup  can give us ready access to XRD data across 20,000 BZ's, the traditional approach for extracting peak widths cannot use such comprehensive information.  Specifically,  the traditional peak width extraction approach of fitting a line cut of high-resolution data is not scalable (see SM section G). This forces researchers to an ad-hoc choice of a handful of peaks,  ruling out statistically meaningful ${\vec{q}\equiv(H,K,L)}$ analysis. Moreover, aiming to identify power law tails of Bragg glass from fitting the peaks is not feasible as the peaks are sharp and span only 2-3 pixels (see SM Fig 5). Instead, we have adopted a high-throughput approach, by combining the automatic \xtec\ extraction of all the CDW peaks and using a new measure of peak width: the peak {\it spread} 
\begin{equation}
\Gamma_{\vec{q}}(T)\equiv\dfrac{I^{\text{Tot}}_{\vec{q}}(T)}{I^{\text{Max}}_{\vec{q}}(T)}\label{EQ_Gamma},
\end{equation}
in units of the number of pixels.
Here, $I^{\text{Tot}}_{\vec{q}}(T)$ is the integrated intensity and $I^{\text{Max}}_{\vec{q}}(T)$ is the maximum intensity (peak height) of the CDW peaks identified by \xtec. The peak spread quantifies how many pixels the peak is effectively spread over [Fig.~2(e)].
While being consistent with the conventional peak width estimates [Fig.~2(f), see section H in SM], 
the spread as defined possesses several merits compared to the traditional extraction of the inverse correlation length.  First,  it is model-independent.  Second,  it does not require high-resolution data. Third,  it naturally integrates with \xtec, which offers the peak boundaries for all the CDW peaks [Fig.~2(e)].  Finally,  when combined with \xtec, $\Gamma_{\vec{q}}(T)$ can reveal the momentum and temperature evolution of peak widths over a statistically significant number of CDW peaks.

Armed with the new high-throughput measure ``peak-spread'' $\Gamma_{\vec{q}}(T)$, we single out the CDW fluctuation contributions by extracting the momentum-independent part of the peak spread by analyzing 
$\Gamma_{\vec{q}}(T)$ across 20,000 BZ's and the entire temperature range [Fig.~3(a-c)]. 
  Specifically, we fit
the momentum dependence of $\Gamma_{\vec{q}}(T)$ at each temperature $T$ to a quadratic function expected in the small $|\vec{q}|$ limit [Fig.~3(b)] ~\cite{Guinier1994,Emery1978PRB,Heilmann1979PRB,Spal1980XraySS,Endres1982JPhys_Chem_Solids}:
\begin{eqnarray}
\Gamma_{\vec{q}\equiv(H,K,L)}(T) = \Gamma_0(T)+ \gamma_H(T) H^2+\gamma_K(T) K^2+\gamma_L(T) L^2,\label{Eq_mom_fit}
\end{eqnarray} 
where $\gamma_H(T)$, $\gamma_K(T)$, and $\gamma_L(T)$  quantify the momentum dependence at each temperature $T$ along $a^*$, $b^*$ and $c^*$ axis respectively. 
In this way, the extracted momentum-independent width $\Gamma_0(T)$ would reflect the peak width solely due to CDW fluctuations (see SM section H for details).  Indeed $\Gamma_0(T)$ extracted from $\sim 3000$ peaks  of the pristine sample drops and plateaus at the critical temperature $T_{\text c1}$[Fig.~3(c)],  as expected for the long-range order signal cut-off by finite experimental resolution (see Fig.~1(d)).

This analysis reveals the emergence of a threshold temperature in intercalated samples shown in Fig.~3(d-f), below which $\Gamma_0(T)$ becomes constant, signifying that the widths of the CDW peaks at these low temperatures are resolution limited. This resolution limit corresponds to an in-plane correlation length of $\sim 20$nm. With the average distance between Pd atoms ($\sim 2$nm for $x=2.9\%$, ~\cite{Straquadine2019Phys.Rev.B}) being smaller than the correlation length, this corresponds to a weak-pinning scenario.  A finite resolution limit is inevitable in any experiment, leaving behind the question of whether the system has long-range order beyond the resolution limit. To go beyond the resolution limit and find the point of vanishing peak width,  we extrapolate $\Gamma_0(T)$  using an empirical formula: 
 \begin{eqnarray}
 \Gamma_0(T)=\overline{\Gamma} +\alpha~(T-\beta)\Theta[T-\beta].\label{EQ_fitting}
\end{eqnarray} 
Here, $\Theta[t>0]=1$ is the Heaviside step function, and $\overline{\Gamma}$ (the resolution limit), $\alpha$, and $\beta$ are the fitting parameters. The linear in $T$ dependence of $\Gamma_0(T)$ is motivated as a first order approximation of $\Gamma_0(T)$  near its vanishing limit. While a series expansion of the width near $T_{\text c1}$ is not strictly valid for the long-range order whose width vanishes as a power law in $T$ near $T_{\text c1}$, the linear approximation will still extrapolate to give the vanishing width with a reasonable value (a lower bound) for $T_{\text c1}$ when applied to temperatures close to $T_{\text c1}$ (Fig 3 (c)). For the intercalated samples, the vanishing trend of the peak spread in Fig~3(d,e) is contrary to the predicted behavior for disorder pinned short-range CDW order where the peak spread increases with lowering temperature~\cite{Maki1986PhysRevBCondensMatter}. On the other hand, it supports a Bragg glass phase, as it is inevitable for the peak spread to vanish below the Bragg glass transition~\cite{Rosso2004Phys.Rev.B}. We estimate the Bragg glass transition temperature to be the positive temperature at which the 
extrapolation reaches vanishing width, i.e., 
$T_{BG_1}=\beta-\bar{\Gamma}/\alpha$. We find positive $T_{BG_1}$ defining 
the Bragg glass phase in all intercalated samples except at the highest concentration [$x=2.9\%$ in Fig.~3(f)]. The $x=2.9\%$ appears to be fluctuating towards Bragg glass without actually reaching Bragg glass. Combining the CDW-1 transition temperature $T_{c1}$ of the pristine sample [from Fig.~2(d)] and the newly extracted Bragg glass transition temperature,  $T_{BG_1}$ of the intercalated samples [Fig.~3(d-f) for 2.0\%, 2.6\% and 2.9\% respectively, and SM Fig.~9(a) for 0.5\%],  we obtain a comprehensive phase diagram shown in Fig.~3(g).  Remarkably, the temperatures identified from the onset of transport anisostropy~\cite{Straquadine2019Phys.Rev.B} align closely with the $T_{BG_1}$, implying that the phase space exhibiting in-plane resistance anisotropy is predominantly covered by the Bragg glass. 

The impact of disorder in the intercalated samples is evidenced by CDW peak asymmetry~\cite{Ravy1993J.Phys.IVFrance,Rosso2003Phys.Rev.B,Ravy2006Phys.Rev.B,Rosso2004Phys.Rev.B} (section D in SM). 
Due to the sharpness of the CDW peaks with their vanishing widths, comparing the peak height between two peaks is prone to pixelation error (section J in SM).  Hence we focus on the asymmetry between the distribution of satellite  scattering intensities across a Bragg peak. As shown in Fig.~4(a-b), the contrast between the raw x-ray data from the pristine sample and the intercalated sample is stark. 
Specifically, while the pristine sample's intensity distribution shows minimal  asymmetry of the satellite diffuse scattering across the Bragg peaks, the intercalated sample clearly shows enhanced asymmetry in the form of half diamonds. The asymmetry is apparent as the congruent satellite points to these half-diamonds show no scattering. The raw intensity cuts shown in Fig.~4(c,d) clearly show that the distribution of satellite intensities across a Bragg peak is uniquely asymmetric but only for the intercalated sample in Fig.~4(d). We quantify the diffuse scattering asymmetry by the ratio $\alpha$ defined as $(I_1-I_2)/(I_1+I_2)$, where $I_1$ and $I_2$ are the respective intensities of the two congruent satellite diamond regions [see Fig~4(e)]. The ratio $\alpha$ in Fig.~4(e) quantifies the enhanced asymmetry in the intercalated samples and its absence in the pristine sample.  The presence of asymmetry  specific to the intercalated sample distinguishes it from the pristine sample and indicates the pinning of modulations in and around the CDW by the intercalant-induced disorder. A comprehensive picture of the prevalent asymmetry in the Bragg glass and short-range ordered phase emerges upon \xtec\ clustering shown in Fig.~4(f-h)  and SM Fig.~9(d). When the entire data set of the intercalated sample is  split into two clusters (after removing the Bragg peaks and the CDW peaks), the clustering results reveal that the diffuse region around satellite CDW-1 peaks is systematically asymmetric.

In summary, we report the first x-ray scattering evidence suggesting the existence of a Bragg glass phase in a family of disordered charge density wave systems, Pd-intercalated ErTe$_3$. In order to disentangle intrinsic phase fluctuation effects of the CDW from crystalline imperfections despite finite experimental resolution, we obtained comprehensive XRD data spanning $\sim 20,000$ Brillouin zones over 30-300K range of temperatures. We then analyzed the entire  $\sim$150GB of XRD data using \xtec, an unsupervised machine learning tool for revealing collective phenomena from voluminous temperature dependent XRD data\cite{Venderley2022Proc.Natl.Acad.Sci.}. We employed a multi-faceted approach of tracking three features, namely the peak height, the peak width, and the asymmetry in satellite diffuse scattering, throughout the entire dataset. Consolidating the results of this analysis, we were able to disentangle the effects of lattice imperfections and finite momentum resolution from the intrinsic tendency for topological quasi-long range order into a Bragg glass phase. We also observe that except for the pristine sample, there is an asymmetry in the satellite diffuse scattering at all intercalation levels, which indicates all intercalated samples have disorder pinning. We thus claim that even an infinitesimal intercalation introduces disorder and  destroys the long-range order to a Bragg glass order. Thus we 
discovered that for $x>0$, the temperature and momentum dependence of the diffraction signal is consistent with the Bragg glass phase spanning most of the phase space that exhibits transport anisotropy, extending up to remarkably high temperatures. Future diffraction experiments at higher resolution and brightness can further confirm the Bragg glass nature of the anisotropic phase by detecting the power-law tails.

The significance of our findings are two-fold. Firstly, we made significant advances in understanding the elusive Bragg glass phase  in a disordered charge density wave system from bulk diffraction data, by mapping the Bragg glass transition temperature estimates $T_{BG}$ in the phase diagram. It is remarkable that the Bragg glass phase suggested from the absence of phase dislocations observed in the STM measurements~\cite{Fang2019Phys.Rev.B} at temperatures below 1.7K extends all the way up to $~100$K and beyond until the Bragg glass phase collapses at around $2.9\%$ intercalation. Moreover, the evidence for the Bragg glass we established leaves only a very narrow range of phase space that can possibly support the competing short-range ordered phase.
Secondly, the new discovery enabled by the use of \xtec\ and the new high-throughput measure of peak-width demonstrates the potential of the new ML-enabled data analysis in addressing fundamental issues when intrinsic fluctuations and the effect of disorder lead to a complex and rich plethora of phenomena. Higher intensity X-ray that can access many BZ's often lead to worse signal-to-noise ratio.
The high-throughput measure of peak width enabled us to disentangle the effects of crystalline defects from the effects of intrinsic CDW phase fluctuations by giving us access to zone-to-zone correlation in fluctuations over 20,000 BZ's. This separation in turn allowed us to connect the voluminous XRD data with the STM observations and the theory of Bragg glasses. The modality of comprehensive high-throughput extraction of theoretically inspired features promises to enable new discoveries in the era of big data, rich with information,  and connect varied facets of complex systems accessible to different probes. 

\section{Acknowledgements} 
This work was supported by the U.S. Department of Energy, Office of Basic Energy Sciences, Division of Materials Sciences and Engineering and used resources of the Advanced Photon Source, a U.S. Department of Energy Office of Science User Facility at Argonne National Laboratory. The initial \xtec\ analysis on the pristine sample was carried out on the Red Cloud at the
Cornell University Center for Advanced Computing, with
the support of the DOE under award DE-SC0018946. The \xtec\ analysis on intercalated samples were carried out on the high powered computing cluster funded in part by the Gordon and Betty Moore Foundation’s EPiQS Initiative, Grant GBMF10436 to E-AK and by the New Frontier Grant  from the college of Arts and Sciences at Cornell to E-AK.  Work at Stanford University (crystal growth and characterization, and contributions to the diffraction experiments) was supported by the Department of Energy, Office of Basic Energy Sciences, under contract DE-AC02-76SF00515. 

\section{Author Contributions} 
J.S., M.D.B., and A.G.S. with supervision by I.R.F. prepared the samples. X-ray experiments and data processing were performed by M.J.K., J.S., R.O., and S.R. K.M. and E.A.K developed and implemented the machine learning analysis, and interpreted the results with guidance and feedback from I.R.F., M.J.K., S.R. and R.O. K.M. and E.A.K. wrote the manuscript with critical inputs from M.J.K., R.O., I.R.F., and all other authors.  

\section{Competing Interests}
The authors declare no competing interests.

\clearpage
\section*{Figures}
\includegraphics[width=0.9\linewidth]{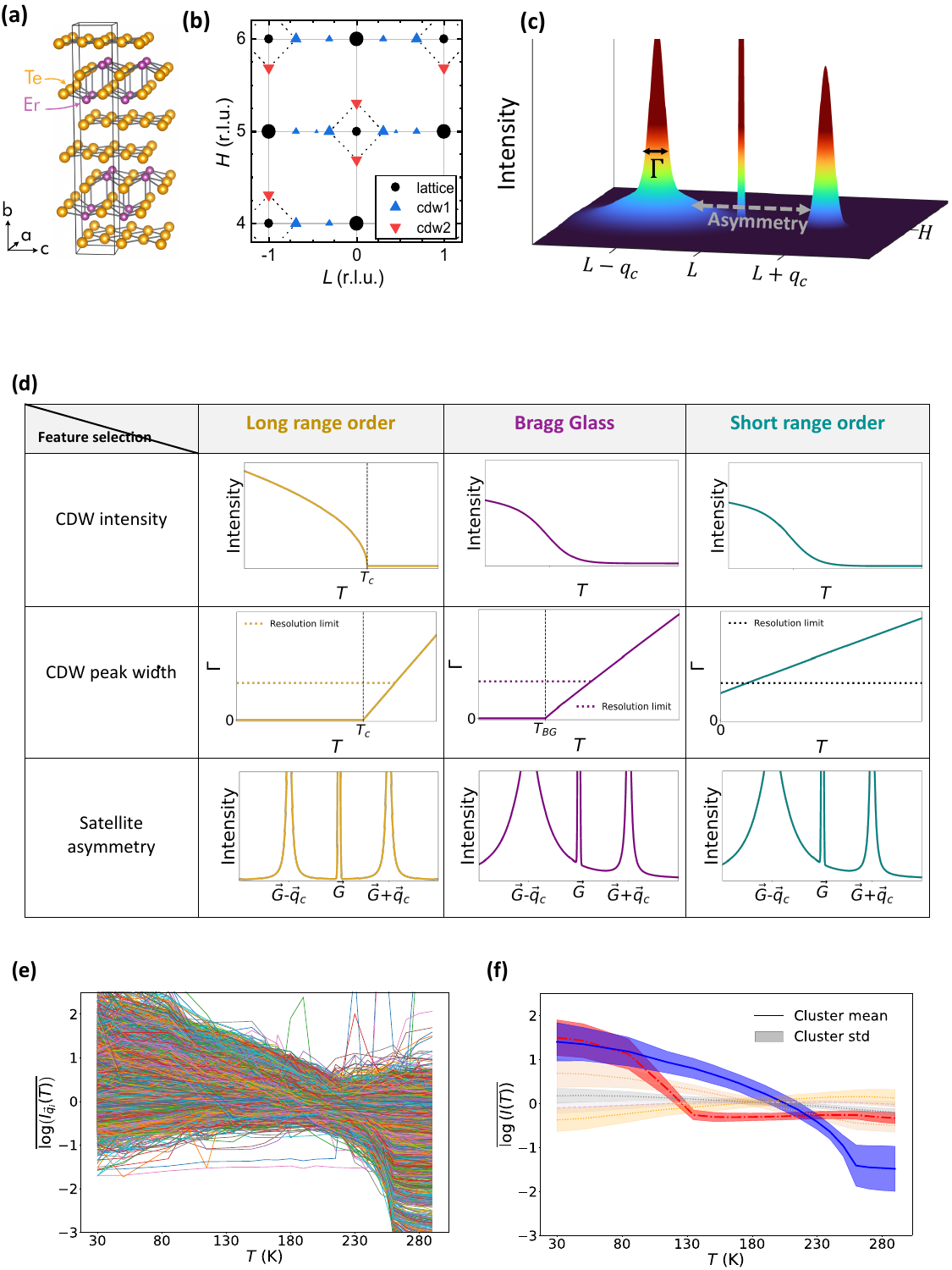}
\ \\
\newpage
\noindent{\bf Fig 1:} Charge density waves in \pdx.
{\bf{(a):}} The crystal structure of pure ErTe$_{3}$. The Te planes have approximately square geometry. The crystal belongs to the $Cmcm$ space group, $b$ denotes the out-of-plane axis, and $a,c$ are the in-plane axes.  {\bf{(b):}} A schematic~\cite{Straquadine2019Phys.Rev.B} showing the Bragg peaks (circles) and CDW peaks (triangles) in the in-plane ($a^*$-$c^*$) reciprocal space. The CDW-1 (up triangle) and CDW-2 (down triangle) satellite peaks are aligned along the $c^*$ and $a^*$ axes, respectively.  {\bf{(c):}} Schematic for the in-plane ($a^*$-$c^*$) intensity distribution of the pair of CDW satellite peaks [at $(H, L\pm q_c)$] around a Bragg peak [at ($H,L$)], with the three features of interest: the intensity of the peak, the width of the peak $\Gamma$ (solid arrow), and the asymmetry in the diffuse scattering surrounding the satellite peaks (dashed arrow).  {\bf{(d):}} Table summarising the diagnostics for classifying the three phases. The first row describes the CDW intensity temperature trajectory. Only the pristine sample with long-range order exhibits a sharp onset, marking the transition temperature $T_{c}$. On the other hand, one cannot distinguish Bragg glass from short-range order as even after the breakdown of Bragg glass order upon increasing temperature, short-ranged fluctuations persist (due to disorder pinning) and contribute to the CDW intensity. The second row illustrates a simplified temperature dependence of the CDW peak width $\Gamma$. The width is zero in the long-range ordered phase below $T_{c}$ of the pristine sample, as well as in the Bragg glass phase below the transition temperature $T_{BG_1}$ of the disordered sample (section B in SM). The vanishing width distinguishes Bragg glass from short-range order. The observed width levels off at the resolution limit (dotted line). The third row illustrates the asymmetry in the diffuse scattering intensity across a Bragg peak. The asymmetry distinguishes pristine (long-range order) from the intercalated sample (Bragg glass and short-range order) as its presence indicates disorder pinning (section C in SM). 
{\bf{(e,f):}} An illustration of \xtec\ to cluster distinct intensity-temperature trajectories, $I(T)$, given the intensity-temperature trajectory  $\{I_{\vec{q}}(T)$ of the pristine ErTe$_3$ sample at various momenta $\vec{q}$ in the reciprocal space. The raw trajectories at each $\vec{q}$ are rescaled as $\overline{\log[I_{\vec{q}}(T)]}=\log[I_{\vec{q}}(T)]-\langle\log[I_{\vec{q}}(T)]\rangle_T$ [panel (e)].  The \xtec\ clusters the trajectories (with color assignments to identify each cluster). The distinct trajectories  $I(T)$ and their standard deviation are shown in panel (f).

\includegraphics[width=\linewidth]{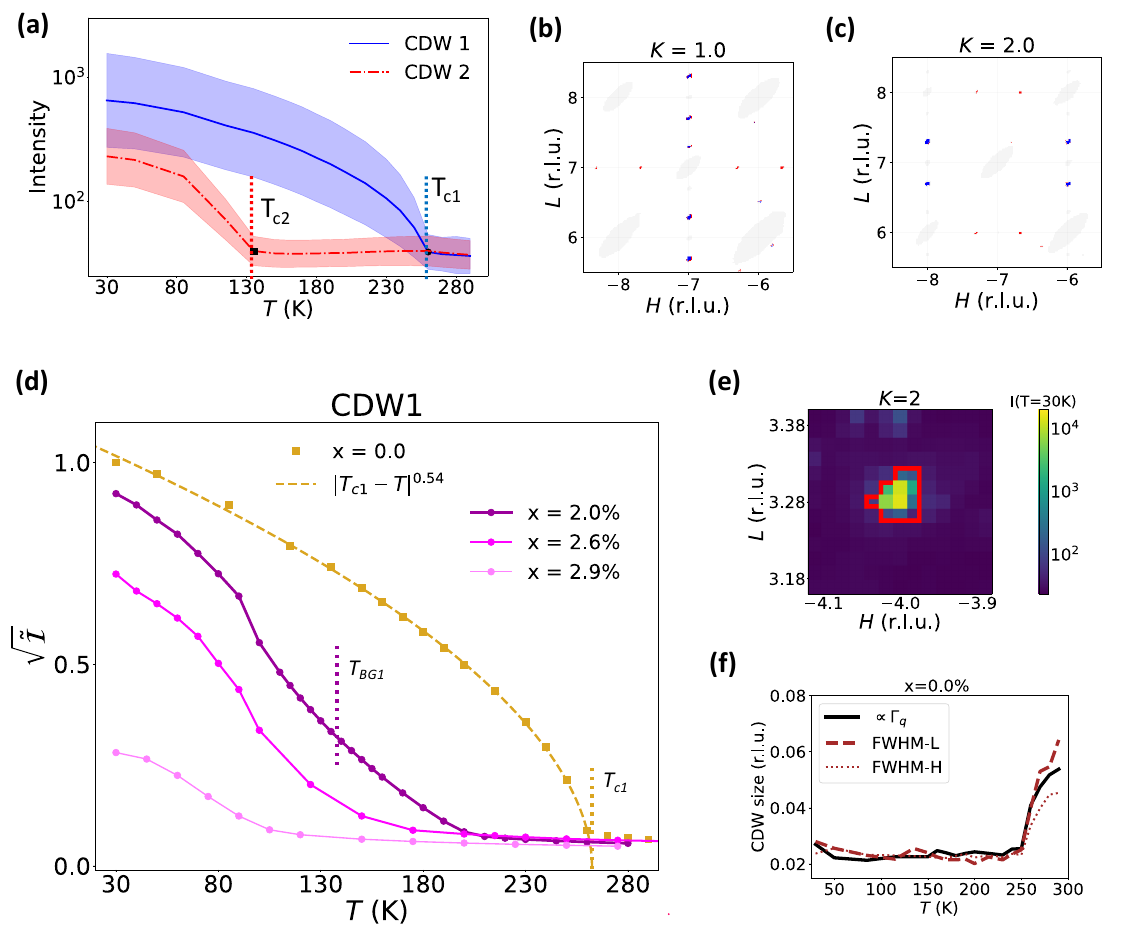}
\newpage
\noindent{\bf Fig 2:} Benchmarking \xtec\ analysis of CDW peak height and peak spread of pristine ErTe$_3$. {\bf{(a):}} \xtec\ reveals the intensity  clusters corresponding to the two CDW order parameter trajectories, color-coded as red and blue. The lines describe the mean and the shaded regions describe one standard deviation of the intensities within each cluster. The estimated transition temperatures $T_{c1}\approx 262K$ for CDW-1 and $T_{c_2}\approx135K$ for CDW-2 are consistent with the temperatures from transport measurements in Ref.~\citenum{Straquadine2019Phys.Rev.B}. {\bf{(b-c):}} A small region of reciprocal space where momenta whose intensity trajectory belongs to the red and blue cluster assignments in (a) are labeled as red and blue pixels, respectively. The red (blue) pixels conform to the CDW-1 (CDW-2) peaks along $c^*$ ($a^*$) axis.  The light grey pixels correspond to Bragg peaks and their diffuse scattering. The 3D structure of the peaks is apparent from the $k=1$ (odd) plane (panel (b)) and $k=2$ (even) plane (panel (c)) that show two different patterns reflecting the $Cmcm$ selection rules governing the Bragg peaks. {\bf{(d):}} The CDW-1 peak averaged intensity (peak height) for \pdx\ at intercalation strength $x=0$, 2.0\%, 2.6\% and 2.9\%. The $\tilde{I}$ is obtained from the average of all the intensities in the CDW-1 cluster ($\sim3000$ peaks), from which we subtract the background intensity contribution. The $\tilde{I}$ for all samples are normalized with the maximum value from $x=0$, for comparison. $\sqrt{\tilde{I}(T)}$ for $x=0$ fits well to a power law $\propto(T_{c1}-T)^{\beta}$ giving  $T_{c1}\sim262$K and $\beta=0.54$ matching the BCS order parameter exponent. The Bragg glass transition temperature $T_{BG_1}$ for $x=2\%$ and $x=2.6\%$ is estimated from the peak width analysis in Fig.~3(e-f).  All solid lines are guides to the eyes. {\bf{(e):}} A CDW-1 peak intensity distribution in the $H$-$L$ plane ($K=2$) for the $x=0\%$ sample at $T=30$K. The red boundary for the CDW-1 peak is estimated by \xtec\ (pixels inside the boundary belong to CDW-1 cluster). Within this boundary, the total intensity $I^{\text{Tot}}_{\vec{q}}(T)$ and maximum intensity $I^{\text{Max}}_{\vec{q}}(T)$ of the peak gives the high throughput measure of peak spread $\Gamma_{\vec{q}}(T)$ [Eq.~\eqref{EQ_Gamma}]. {\bf{(f):}} The peak spread  ($\Gamma$) of the CDW peak in (e), along with the FWHM from line cuts along $H$ (FWHM-H) and $L$ (FWHM-L), at various $T$ for $x=0\%$.

\newpage

\includegraphics[width=\linewidth]{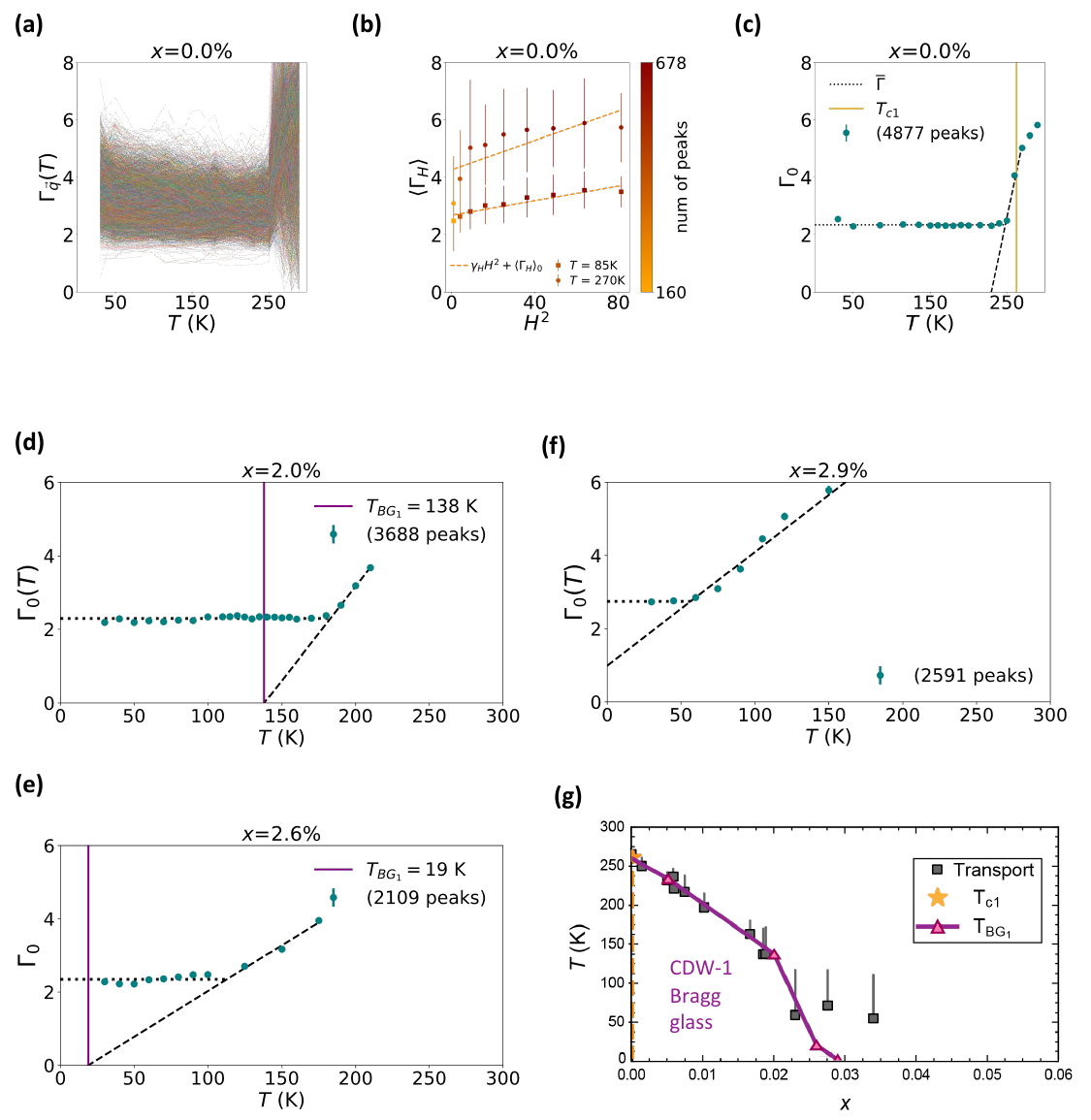}
\newpage
\noindent{\bf Fig 3:} Momentum independent peak spread and Bragg glass transition. {\bf{(a):}} Peak spread $\Gamma_{\vec{q}}(T)$ of all  CDW-1 peaks in the $x=0$ data. {\bf{(b):}} The quadratic momentum ($H^2$) dependence of $\Gamma_{\vec{q}}(T)$. $\langle\Gamma_{H}\rangle$ (symbols) is obtained by averaging $\Gamma_{\vec{q}}$ over values of $K$ and $L$ that share the same $|H|$. The symbols show the mean $\pm$ one standard deviation. The markers are color-coded, and the color bar indicates the number of peaks determining the statistics of each marker. {\bf{(c):}} From the erratic and broad distribution of $\Gamma_{\vec{q}}(T)$ in panel (a), the momentum-independent spread $\Gamma_0$ extracted from the 3D  quadratic fit [Eq.~\eqref{Eq_mom_fit}] shows a $T$ independent (resolution limited) spread below $T_{c1}$, where $T_{c1}=262$ K from Fig.~2 (d). Symbols give the fit estimate, and error bars give 95\% confidence bounds on $\Gamma_0$ estimated from 4877 peaks. The dashed line shows the linear extrapolation [Eq.~\eqref{EQ_fitting}] to vanishing width. {\bf{(d-f):}} The $\vec{q}$ independent broadening of CDW-1 peak spread, $\Gamma_0(T)$, extracted by fitting their $\Gamma_{\vec{q}}$ to a quadratic function of $\vec{q}$ [Eq.~\eqref{Eq_mom_fit}] for $x=2\%$, 2.6\%, and 2.9\% in panels (d), (e) and (f) respectively. Symbols give the fit estimate, and error bars give 95\% confidence bounds on $\Gamma_0$ estimated from 3688 (2.0\%), 2109 (2.6\%) and 2591 (2.9\%) peaks. Dashed lines are a phenomenological fitting function [Eq.~\eqref{EQ_fitting}] to extract CDW-1 Bragg glass temperature $T_{BG_1}$. The dotted lines mark the resolution limit $\overline{\Gamma}$ from the fit. We find a Bragg glass regime for $x=2\%$ and 2.6\% sample by extracting $T_{BG_1}$ (vertical solid lines) from extrapolating the broadening regime to zero spread. {\bf{(g)}:}  Our estimates for the transition temperatures $T_{c1}$ of $x=0\%$ (star symbol) and $T_{BG_1}$ of $x>0$ (up triangle symbols)   are overlaid on the phase diagram from the in-plane resistance anisotropy measurements  (square symbols) from Ref.~\citenum{Straquadine2019Phys.Rev.B}. Lines are guides to the eyes. The CDW-1 long range ordered phase of $x=0\%$ is indicated by the dashed orange line, and the CDW-1 Bragg glass phase lies below $T_{BG_1}$ for $x>0$.

\includegraphics[width=\linewidth]{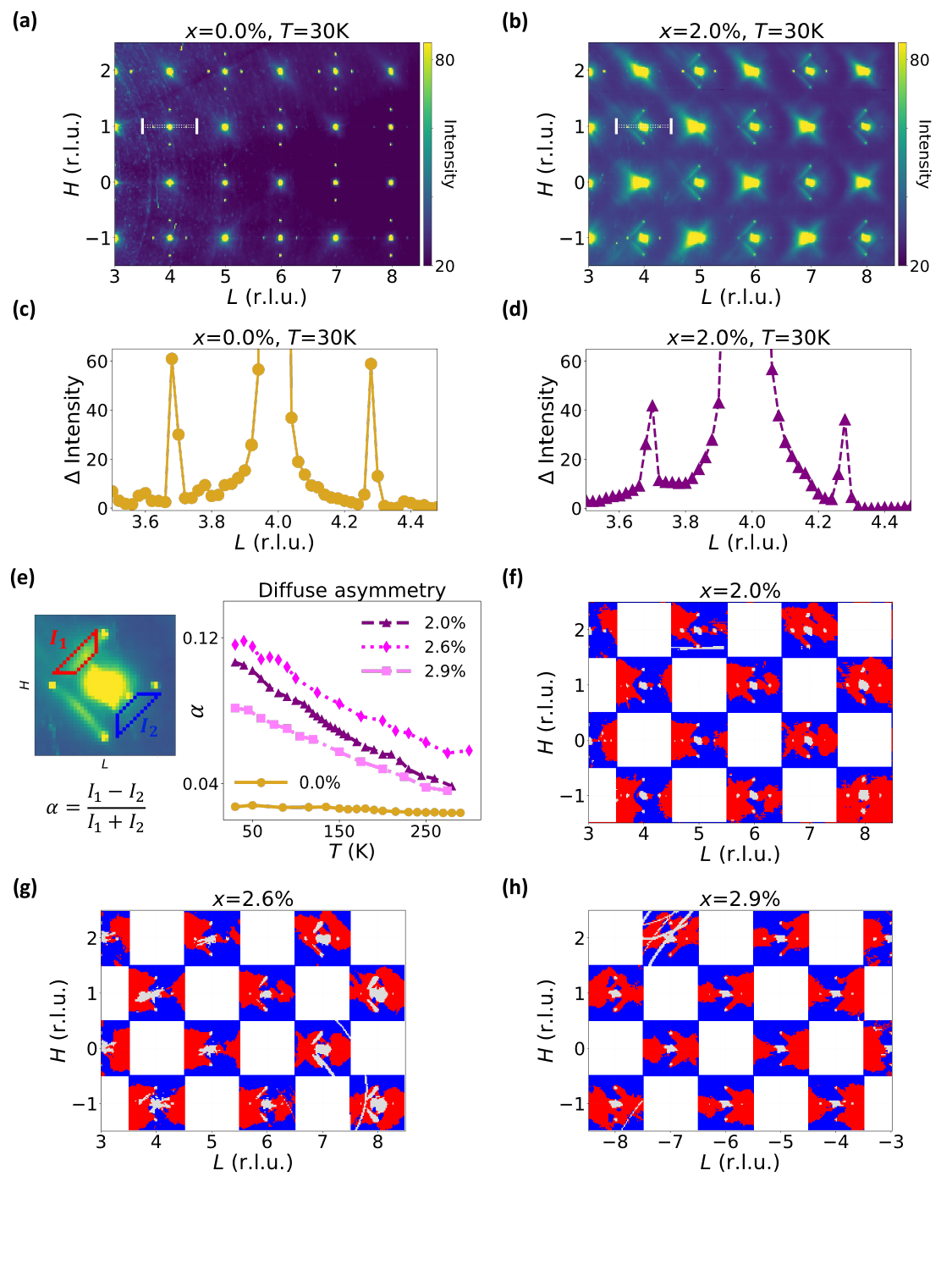}
\newpage
\noindent{\bf Fig 4:} Intensity asymmetry of the CDW satellite peaks. {\bf{(a,b)}:} The x-ray intensity at $T=30$K, in the $H$-$L$ plane with $K$ (out-of-plane axis) averaged over all  values ($-20\le K\le20$), for the pristine sample [$x=0\%$ in panel(a)] and intercalated sample [$x=2\%$ in panel (b)]. Only the intercalated sample shows a diffuse scattering that is asymmetrically distributed between the two satellite peaks, in the form of half diamonds. The white horizontal lines across a pair of CDW-1 satellite peaks mark the region along which a line cut is taken and is shown in panels (c-d). {\bf{(c,d)}:} Line cut intensities $I(L)$ for the pristine  (panel c) and 2\% intercalated sample (panel d) at 30K. Line cuts are along $3.5\le L\le4.5$ (r.l.u) with intensity averaged over $H\in [0.98,1.02]$ (r.l.u) and $K\in[-20,20]$.  {\bf{(e)}:} Asymmetry in diffuse scattering, quantified by the ratio $\alpha=(I_1-I_2)/(I_1+I_2)$ for each satellite pair, where $I_1$ and $I_2$ are the intensities averaged within the upper left arm (red boundary) and the lower right arm (blue boundary) respectively. The plot shows the average value of $\alpha$ evaluated for the satellite diffuse diamonds of all the $H$+$L$ = odd Bragg peaks in the reciprocal space spanned by panels (a) and (b). {\bf{(f,g,h)}:} Two cluster \xtec\ results color coded as red and blue, from the temperature trajectories  of the diffuse scattering intensities  of the $x=2\%$ intercalated sample (f), 2.6\% (g) and 2.9\% (h). The  asymmetric distribution of red and blue clusters surrounding the CDW satellite peaks is systematically present  in the three intercalated samples, clearly revealing the signature of disorder pinning. The intensities of the CDW peaks and $H$+$L$ = odd Bragg peaks  (white pixels, identified from a prior \xtec \ analysis)  are excluded from this two-cluster \xtec, along with the $H$+$L$ = even Bragg peaks removed by a square mask (square white regions).

\newpage

\section{Methods}

\textbf{Synthesis:}  Samples were grown using a Te self-flux method as described in Ref.~\citenum{Ru2006Phys.Rev.B}. Small amounts of Pd were included in the melt to produce the palladium intercalated crystals. Crystals produced had an area of 1-2mm across and varied in thickness with intercalation level. Since the CDW transition temperature is well characterized for different intercalation levels, resistivity measurements of the sample batches used were taken to determine the intercalation levels of the samples studied~\cite{Straquadine2019Phys.Rev.B}. 

\textbf{X-ray scattering:} Samples were shipped to Argonne in sealed vials filled with inert gas and removed and mounted on the tips of polyimide capillaries just before measurement to avoid degradation from water and oxygen exposure. During measurements, samples were cooled using an Oxford N-Helix Cryostream, which surrounded samples with either N$_2$ or He gas. Measurements were taken with incident x-ray energy of 87 keV in transmission geometry, with samples continuously rotated at 1$\degree$ s$^{-1}$ and a Pilatus 2M CdTe detector taking images at 10 Hz. For each sample at each temperature, three such 365$\degree$ rotation scans were collected, with the detector slightly offset and the rotation angle slightly changed to fill in detector gaps and allow for removal of detector artifacts (detailed in Ref. \citenum{Krogstad2019tc}).

{\bf Data analysis:}  The \xtec\ package used for the analysis can be installed through the PyPI distribution, or from the source \href{https://github.com/KimGroup/XTEC}{github.com/KimGroup/XTEC}. Further details regarding the data analysis are provided in the supplementary material (SM).  

\section{Data and Code Availability:}
All the codes used for the analysis are available at \href{https://github.com/KimGroup/PdxErTe3_XTEC_analysis}{github.com/KimGroup/PdxErTe3\_XTEC\_analysis}. A dataset containing $\sim$2000 CDW peak features extracted by \xtec\ for each intercalated sample, and scripts to analyze them are available at 
Figshare  \href{https://figshare.com/s/3058d505bfffed3a7436}{https://figshare.com/s/3058d505bfffed3a7436}. Any data not deposited online will be shared with interested researchers upon request.

\clearpage


\newpage
\bibliographystyle{apsrev4-2}
\bibliography{PdRTe3_refs}

\clearpage

\pagebreak
\clearpage
\appendix
\onecolumngrid
\renewcommand{\thefigure}{S\arabic{figure}}
\renewcommand{\thetable}{S\arabic{table}}

\section{Supplementary Material}

\section{A: Scaling argument for Bragg glass}\label{SMA}
 The energy scaling argument by Imry and Ma considering an XY model~\cite{Imry1975Phys.Rev.Lett.}, and by Fukuyama and Lee considering a disordered CDW~\cite{Fukuyama1978Phys.Rev.Ba}, indicates that only short ranged correlations are allowed in continuous symmetry broken states below 4 dimensions, based on an assumption of a disorder potential linearly coupled to the phase. However for the CDW, the true potential is non-linear and a periodic function of phase.   Nattermann~\cite{Nattermann1990Scaling} took this periodicity into account and showed that the modified scaling argument supports the quasi-long range order of the Bragg glass. Here we recall the scaling arguments, starting with Imry and Ma's analysis and its shortcoming, and then follow Nattermann's analysis  [Ref.~\citenum{Nattermann1990Scaling}] supporting the Bragg glass order in 3D.  We start with a charge density wave, 
\begin{eqnarray}
\rho(\vec{r},\phi)=\rho_0 \cos[\vec{q}_c\cdot\vec{r}+\phi(\vec{r})]
\end{eqnarray}
with  an incommensurate wave vector $\vec{q}_c$, a constant amplitude $\rho_0$ and a phase $\phi(\vec{r})$ that can spatially vary due to thermal fluctuations and disorder interactions. The interaction with quenched disorder in $D$ spatial dimensions can be described with an elastic model whose Hamiltonian is given by~\cite{Rosso2004Phys.Rev.B},
\begin{eqnarray}
H=\frac{C}{2}\int d^D r~|\vec{\nabla}\phi(\vec{r})|^2+ V_0\int d^D r ~\Sigma(\vec{r})\rho(\vec{r},\phi) \label{Eq:elastic_H}
\end{eqnarray}
where the first term is the elastic part with $C$ as the elastic stiffness, and the second term is the disorder potential due to quenched impurities exerting a potential $V_0$ on the charge density, and distributed with a probability density $\Sigma(\vec{r})$. We assume there are no topological defects in the system so that $\phi(\vec{r})$ is single valued and the elastic model is well defined, which is a necessary condition for a Bragg glass~\cite{Giamarchi1995Phys.Rev.B,Gingras1996Phys.Rev.B,Giamarchi1997Phys.Rev.B}.  A spatially modulated phase $\phi(\vec{r})$ increases the elastic energy, but can lower the potential energy by conforming $\rho(\vec{r})$ to the impurity distribution. The disordered phases arise from this competition between the elastic energy cost and the potential energy gain.
These phases are distinguished by the fluctuations in $\phi(\vec{r})$ relative to an arbitrary reference point $\phi(\vec{r}=0)$, given by
\begin{eqnarray}
W^2(|\vec{r}|)&=&\langle\overline{\left(\phi(\vec{r})-\phi(0)\right)^2}\rangle \label{Eq:W(r)}
\end{eqnarray}
where $\langle\dots\rangle$ denotes a thermal average and $\overline{(\dots)}$ denotes a disorder ensemble average. To simplify, we fix $\phi(0)=0$, and assume that fluctuations are spherically symmetric with respect to $\vec{r}=0$. 

We first identify the scaling of elastic energy cost from Eq.~\eqref{Eq:elastic_H}. For a phase that varies by an amount $W(R)$ over a distance $R$, the elastic energy (EE) in the volume $R^D$ scales as 
\begin{eqnarray}
\text{EE} \propto \frac{1}{2}C\left(\dfrac{W(R)}{R}\right)^2 R^{D} \label{Eq:elastic_cost}
\end{eqnarray}
This shows that for $D>2$, the elastic energy cost increases with phase fluctuations over larger distances. In the absence of disorder, this energy cost protects the long range order.

Now we discuss the scaling of the potential energy in a volume $R^D$. We imagine each site is independently occupied by an impurity with probability $n_I$, the impurity concentration. The volume includes $n_IR^D$ impurities,  and each random impurity site $\vec{r}_i$ contributes a potential energy $V(\vec{r}_i)$ given by,
\begin{eqnarray}
V(\vec{r}_i)= V_0\rho_0\cos[\vec{q}_c.\vec{r}_i+\phi(\vec{r}_i)] \label{Eq:dis_potential}
\end{eqnarray}

\underline{\textbf{Imry-Ma scaling}}:
Imry and Ma's argument is valid when  $\phi(\vec{r}_i)$ is small and a linear approximation applies to  Eq.~\eqref{Eq:dis_potential}, given by
\begin{eqnarray}
V(\vec{r}_i)=V_0\rho_0\left( \cos(\vec{q}_c\cdot\vec{r}_i)-\sin(\vec{q}_c\cdot\vec{r}_i)\phi(\vec{r}_i)\right) +\mathcal{O}(\phi^2(\vec{r}_i))\label{Eq:linear_dis}
\end{eqnarray}
where we can discard the first term that sets a constant offset, and the second term gives the potential energy gain from $\phi(\vec{r}_i)$. To estimate the magnitude of this energy gain in a volume $R^D$, we note that a typical impurity site has a position $|\vec{r}_i|\sim R$ and the phase $|\phi(\vec{r}_i)|\approx \left(\overline{\langle\phi^2(\vec{r}_i)\rangle}\right)^{1/2}\sim W(R)$. Hence, the magnitude of potential energy gain from each impurity, $V_0\rho_0\left|\sin(\vec{q}_c\cdot\vec{r}_i)\phi(\vec{r}_i)\right|$, has a typical value $\sim V_0\rho_0W(R)$, and the magnitude of total potential energy (PE) scales as  
\begin{eqnarray}
\text{PE}&\sim& \left(\sqrt{n_IR^D}\right) V_0\rho_0W(R) \label{Eq:Imry_pot_gain}
\end{eqnarray}
where the factor $\sqrt{n_IR^D}$ follows from central limit theorem giving the root mean squared value from $n_IR^D$ independent random impurities.

Equating the elastic energy cost [Eq.~\eqref{Eq:elastic_cost}] to potential energy gain [Eq.~\eqref{Eq:Imry_pot_gain}] gives the optimal $W(R)$ given by
\begin{eqnarray}
CR^{D-2}W^2(R) &\sim& \left(\sqrt{n_I R^{D}}\right) V_0\rho_0 W(R)\\
\Rightarrow W(R)&\sim&  \left(\dfrac{V_0\rho_0n_I^{1/2}}{C}\right) R^{(4-D)/2} \label{Eq:Imry_W}
\end{eqnarray}
For $D<4$, $W(R)$ grows algebraically with $R$, tempting one to conclude that the system is short-range-ordered for arbitrarily small disorder strength. A length scale for the short range order was estimated as the length $R_0$ at which $W(R_0)\sim \pi$, the maximum value for the fluctuation. From Eq.~\eqref{Eq:Imry_W}, an estimate for this length scale $R_0$ (also known as the Fukuyama-Lee length ~\cite{Fukuyama1978Phys.Rev.Ba}) is given by
\begin{eqnarray}
R_0=\left(\dfrac{ C}{V_0\rho_0n_I^{1/2}}\right)^{2/(4-D)} \label{Eq:Imry_length}
\end{eqnarray}
$R_0$ is also a length scale that highlights the breakdown of the above scaling argument. At these length scales, $\phi(\vec{r})$ is large and inconsistent with the linear approximation in Eq.~\eqref{Eq:linear_dis}. The full periodic nature of the potential needs to be considered to understand the fluctuations beyond $R_0$.

\underline{\textbf{Nattermann's scaling}}: Retaining the periodic nature of the potential energy in Eq.~\eqref{Eq:dis_potential}, we can now get the new scaling estimate for the magnitude of potential energy in a volume $R^D$ as follows. Each impurity contributes to the potential energy by a  magnitude $V_0\rho_0\left|\cos(\vec{q}_c.\vec{r}_i+\phi(\vec{r}_i))\right|\sim V_0\rho_0e^{-\overline{\langle\phi^2(\vec{r}_i)\rangle}/2}$. In a volume $R^D$,  since the typical position $|\vec{r}_i|\sim R$,  $V_0\rho_0e^{-\overline{\langle\phi^2(\vec{r}_i)\rangle}/2}\sim V_0\rho_0e^{-W^2(R)/2}$, and the total potential energy (PE) thus scales as 
\begin{eqnarray}
\text{PE}&\sim& \left(\sqrt{n_IR^D}\right)V_0\rho_0 e^{-W^2(R)/2} \label{Eq:pot_gain}
\end{eqnarray}
where the factor $\sqrt{n_IR^D}$ follows from fluctuations of $n_IR^D$ independent random impurities.

Equating the elastic energy cost [Eq.~\eqref{Eq:elastic_cost}] to potential energy gain [Eq.~\eqref{Eq:pot_gain}] gives the optimal $W(R)$ given by
\begin{eqnarray}
CR^{D-2}W^2(R) &\sim& \left(\sqrt{n_IR^D}\right)V_0\rho_0 e^{-W^2(R)/2}\\
\Rightarrow W^2(R)&\sim&  (4-D)\log(R/R_0) +\mathcal{O}\left(\log(\log(R/R_0))\right)
\end{eqnarray}
where $R_0$ is the same length scale from Eq.~\eqref{Eq:Imry_length}. Thus for $D<4$, $W^2(R> R_0)$ grows logarithmically to leading order. This is the Bragg glass order.

\section{B: Sample preparation and X-ray details}\label{SMB}

Samples were grown using a Te self-flux method as described in Ref.~\citenum{Ru2006Phys.Rev.B}. Small amounts of Pd were included in the melt to produce the palladium intercalated crystals. Crystals produced had an area of 1-2mm across and varied in thickness with intercalation level. Since the CDW transition temperature is well characterized for different intercalation levels, resistivity measurements of the sample batches used were taken to determine the intercalation levels of the samples studied~\cite{Straquadine2019Phys.Rev.B}.  Samples were shipped to Argonne in sealed vials filled with inert gas and removed and mounted on the tips of polyimide capillaries just before measurement to avoid degradation from water and oxygen exposure. During measurements, samples were cooled using an Oxford N-Helix Cryostream, which surrounded samples with either N$_2$ or He gas. Measurements were taken with incident x-ray energy of 87 keV in transmission geometry, with samples continuously rotated at 1$\degree$ s$^{-1}$ and a Pilatus 2M CdTe detector taking images at 10 Hz. For each sample at each temperature, three such 365$\degree$ rotation scans were collected, with the detector slightly offset and the rotation angle slightly changed to fill in detector gaps and allow for removal of detector artifacts (detailed in Ref. \citenum{Krogstad:2019tc}).

\section{C: Peak width of a disordered CDW}\label{SMC}

We describe the relationship between the  CDW peak width and the density correlations in a Bragg glass and short-range-ordered phase, following the analysis from Refs.~\cite{Ravy2006Phys.Rev.B,Rosso2003Phys.Rev.B,Rosso2004Phys.Rev.B}. Consider a 3D lattice with $N$ sites, and atoms arranged at  $\vec{r}_n=\vec{R}_n+\vec{c}_n$ where $\vec{R}_n$ are the  crystal lattice positions, and $\vec{c}_n$ are the lattice displacements due to a CDW. Let us describe the lattice displacements due to a unidirectional CDW with an incommensurate modulation vector $\vec{q}_c$, given by
\begin{eqnarray}
\vec{c}_n=\vec{c}_0\cos(\vec{q}_c\cdot\vec{R}_n+\phi_n)
\end{eqnarray}
 where $\phi_n$ is a non uniform phase with  fluctuations due to disorder interaction, and $\vec{c}_0$ is a uniform amplitude (amplitude fluctuations are energetically more expensive, hence neglected). The scattering intensity at a momentum $\vec{Q}$ is given by 
\begin{eqnarray}
I(\vec{Q})&=&\sum_{n,m}e^{i\vec{Q}\cdot(\vec{R}_n-\vec{R}_m)}\langle e^{i\vec{Q}\cdot(\vec{c}_n-\vec{c}_m)}\rangle_{_{\phi}}\label{Eq_IQ_phase}
\end{eqnarray}
where $\langle\cdots\rangle_{_{\phi}}$ denotes  ensemble average over disordered phase configurations $\{\phi_n\}$, and we assume a uniform disorder averaged form factor set to unity for all atoms. For small $\vec{c}_0$, the $I(\vec{Q})$ is simplified to,
\begin{eqnarray}
I(\vec{Q})&=&\sum_{n,m}e^{i\vec{Q}\cdot(\vec{R}_n-\vec{R}_m)}\left[1-\dfrac{1}{2}\langle\left(\vec{Q}\cdot(\vec{c}_n-\vec{c}_m)\right)^2\rangle_{_{\phi}}\right]+\mathcal{O}(|\vec{Q}\cdot\vec{c}_0|^4)\nonumber\\
&\approx&\sum_{n,m}e^{i\vec{Q}\cdot(\vec{R}_n-\vec{R}_m)}\left[\left(1-\dfrac{1}{2}(\vec{Q}\cdot\vec{c}_0)^2\right)+\dfrac{1}{4}(\vec{Q}\cdot\vec{c}_0)^2\left(e^{i\vec{q}_c\cdot(\vec{R}_n-\vec{R}_m)}\langle e^{i(\phi_n-\phi_m)}\rangle_{_{\phi}}+e^{-i\vec{q}_c\cdot(\vec{R}_n-\vec{R}_m)}\langle e^{-i(\phi_n-\phi_m)}\rangle_{_{\phi}}\right)\right]\nonumber
\label{Eq_Iq_dis}
\end{eqnarray}
From the above expression, we can deduce the two CDW satellite peaks at $\vec{Q}=\vec{G}\pm \vec{q}_c$ around each Bragg peak at $\vec{G}$. Focusing on the satellite peak  around $\vec{G}+ \vec{q}_c$, the intensity profile is given by
\begin{eqnarray}
I(\vec{Q}=\vec{G}+ \vec{q}_c+\delta\vec{q})&=&\dfrac{1}{4}(\vec{Q}\cdot\vec{c}_0)^2\sum_{n,m}e^{i\delta\vec{q}\cdot(\vec{R}_n-\vec{R}_m)}\langle e^{-i(\phi_n-\phi_m)}\rangle_{_{\phi}} \label{Eq_Ic_sum}
\end{eqnarray}
where $|\delta\vec{q}|\ll|\vec{q}_c|$. The density correlations $\langle e^{-i(\phi_n-\phi_m)}\rangle_{_{\phi}}$, which using the Gaussian approximation for small fluctuations get simplified to, 
\begin{eqnarray}
\langle e^{-i(\phi_n-\phi_m)}\rangle_{_{\phi}}&=&   e^{-\frac{1}{2}\langle(\phi_n-\phi_m)^2\rangle_{_{\phi}}} + \mathcal{O}[\langle(\phi_n-\phi_m)^4\rangle_{_{\phi}}].\label{Eq_gauss_approx}
\end{eqnarray}
Due to translational symmetry of the disorder averaged phase fluctuations, we can define the density correlation function in terms of  fluctuations relative to a reference point, given by
\begin{eqnarray}
C_{\phi}(\vec{r})=e^{-\frac{1}{2}\langle(\phi(\vec{r})-\phi(0))^2\rangle_{_{\phi}}},\label{Eq_Cr}
\end{eqnarray} 
where $\phi(\vec{r}=\vec{R}_n)\equiv\phi_n$. Substituting Eq.~\eqref{Eq_gauss_approx} and \eqref{Eq_Cr} in Eq.~\eqref{Eq_Ic_sum},  we get the CDW satellite intensity as
\begin{eqnarray}
I(\vec{Q}=\vec{G}+ \vec{q}_c+\delta\vec{q})\approx \dfrac{1}{4}(\vec{Q}\cdot\vec{c}_0)^2N\mathit{v}^{-1}\int (d^3\vec{r}) e^{i\delta\vec{q}\cdot\vec{r}}C_{\phi}(\vec{r})\label{Eq_Ic_integral}
\end{eqnarray}
where we have replaced the discrete lattice sum with an integral over $\vec{r}\equiv\vec{R}_n-\vec{R}_m$, and $\mathit{v}^{-1}$ is the volume of a unit cell. The profile of the CDW peak is thus determined by  $C_{\phi}(\vec{r})$,  whose  long distance behavior distinguishes long-range-ordered, Bragg glass, and short-range-ordered CDW phases. 

\textbf{1. Long range ordered CDW}: $C_{\phi}(\vec{r}\rightarrow \infty)\ne 0$ for a CDW with perfect long range ordered phase. 
Here, Eq.~\eqref{Eq_Ic_integral} gives  delta function peaks with ideally zero peak width.  

\textbf{2. Short range ordered CDW}: When  $C_{\phi}(\vec{r})\sim e^{-r/\zeta}$, with a correlation length  $\zeta$, Eq.~\eqref{Eq_Ic_integral}   gives a broadened (nearly Lorentzian) peak at $\vec{Q}=\vec{G}\pm \vec{q}_c+\delta\vec{q}$ given by
\begin{eqnarray}
I(\vec{Q})\propto (\vec{Q}\cdot\vec{c}_0)^2\zeta^3 \dfrac{1}{\left(1+\zeta^2|\delta\vec{q}|^2\right)^2}
\end{eqnarray}
whose full width at half maxima (FWHM) is $(2\sqrt{\sqrt{2}-1})\zeta^{-1}$. Thus the observed peak width is determined by the inverse phase correlation length $\zeta^{-1}$, and is independent of the momentum $\vec{Q}$ of the peak.

\textbf{3. Bragg glass ordered CDW}: A Bragg glass phase is distinguished by a power law decaying phase correlation: $C_{\phi}(r > R_0)\sim \left(r/R_0\right)^{-\eta}$ where $\eta\approx 1$ in 3D is a universal exponent as shown by Refs.~\cite{Giamarchi1994PRL,Giamarchi1995Phys.Rev.B}, and $R_0$ is a small distance cut-off [see Eq.~\ref{Eq:Imry_length}]  that sets the onset of power law decay.  For the Bragg glass, Eq.~\eqref{Eq_Ic_integral} in the limit $|\delta\vec{q}|\rightarrow 0$ can be solved to get the intensity at $\vec{Q}=\vec{G}\pm \vec{q}_c+\delta\vec{q}$ as 
\begin{eqnarray}
I(\vec{Q})
&\propto& \left(|\delta\vec{q}|^{\eta-3}\right)(\vec{Q}\cdot\vec{c}_0)^2R_0^{\eta} 
\end{eqnarray}
For 3D where $\eta=1$, the peak intensity of a Bragg glass diverges as $|\delta\vec{q}|^{-2}$. As with long range order, the observed width will be the resolution limit of the detector~\cite{Rosso2004Phys.Rev.B}.

\section{D: Disorder pinning and asymmetry}\label{SMD}

Here we show that the presence of an asymmetry between the satellite peak intensities signals the disorder pinning of lattice modulations. The  derivation below follows from Refs~\cite{Guinier1994,Ravy2006Phys.Rev.B}. While the asymmetry  signature was  experimentally observed for short range ordered CDW materials  ~\cite{ravy1992destructive,Ravy1993J.Phys.IVFrance,Ravy2006Phys.Rev.B}, they were also  predicted to occur in Bragg glass ordered CDW in Ref.~\cite{Rosso2003Phys.Rev.B,Rosso2004Phys.Rev.B}.

Consider a 3D lattice with atoms arranged at  $\vec{r}_n=\vec{R}_n+\vec{u}_n$ where $\vec{R}_n$ are the  crystal lattice positions and $\vec{u}_n$ is a displacement from the $n^{\text{th}}$ lattice site. The Fourier component of the displacement modulation is given by
\begin{eqnarray}
\vec{u}_{\vec{q}}=N^{-1/2}\sum_n \vec{u}_{n}e^{-i\vec{q}\cdot\vec{R}_n} \label{Eq_uq}
\end{eqnarray}
where $N$ is the total number of sites. Let us model the intercalation (disorder) as modifying the original form factor to a new value $f_j$ at random sites $j$. The Fourier component of the modulated form factor is given by 
\begin{eqnarray}
\tilde{f}_{\vec{q}}=N^{-1/2}\sum_n f_{n}e^{-i\vec{q}\cdot\vec{R}_n}. \label{Eq_fq}
\end{eqnarray}
The scattering intensity at a momentum $\mathcal{Q}$ for this model with intercalation disorder and small lattice displacements is given by
\begin{eqnarray}
I(\vec{Q})&=&\sum_{n,m}e^{i\vec{Q}\cdot(\vec{R}_n-\vec{R}_m)}\langle f_nf_m e^{i\vec{Q}\cdot(\vec{u}_n-\vec{u}_m)}\rangle\\
&=&\sum_{n,m}e^{i\vec{Q}\cdot(\vec{R}_n-\vec{R}_m)}\langle f_nf_m \left[1+i\vec{Q}\cdot(\vec{u}_n-\vec{u}_m)\right]\rangle + \mathcal{O}\left(|\vec{Q}\cdot(\vec{u}_n-\vec{u}_m)|^2\right)\label{Eq_Iq}
\end{eqnarray}
where $\langle\cdots\rangle$ denotes thermal and disorder average. 

We are interested in the asymmetry of the intensities $I(\vec{Q})$ between the two satellite points $\vec{G}\pm \vec{q}$ across a Bragg peak at $\vec{G}$, where $\vec{q}$ is within the first Brillouin zone. Substituting the inverse Fourier transforms of Eq.~\eqref{Eq_uq} for $\vec{u}_i$ and Eq.~\eqref{Eq_fq} for $f_i$ in to Eq.~\eqref{Eq_Iq}, we get the satellite asymmetry to be 
\begin{eqnarray}
I(\vec{G}+\vec{q})-I(\vec{G}-\vec{q})=2i\tilde{f}_{0}\vec{G}\cdot\left(\langle\vec{u}_{-\vec{q}}\tilde{f}_{\vec{q}}\rangle-\langle\vec{u}_{\vec{q}}\tilde{f}_{-\vec{q}}\rangle\right)+\mathcal{O}\left(|\vec{G}\cdot\vec{u}_{\vec{q}}|^2\right) \label{Eq_asymmetry}
\end{eqnarray}
where $N^{-1/2}\tilde{f}_{0}=N^{-1}\sum_j f_{j}$ is the average form factor of the disordered lattice. If the lattice displacement modulations are not correlated with the intercalant positions (no disorder pinning), then the term $\langle\vec{u}_{\vec{q}}\tilde{f}_{-\vec{q}}\rangle=\langle\vec{u}_{\vec{q}}\rangle\langle\tilde{f}_{-\vec{q}}\rangle=0$ since $\langle\vec{u}_{\vec{q}}\rangle=\langle\vec{u}_{\vec{-q}}\rangle=0$. The $\langle\vec{u}_{\vec{q}}\rangle=0$ is true for both incommensurate long range ordered CDW (since the CDW phase in each disorder configuration is arbitrary) and for short range ordered displacements (the disorder average of the displacements is zero).  Thus the leading order contribution to the intensity asymmetry is zero in the absence of disorder pinning of lattice modulations.

On the other hand, in the presence of disorder pinning,  $\langle\vec{u}_{\vec{q}}\tilde{f}_{-\vec{q}}\rangle\ne\langle\vec{u}_{\vec{q}}\rangle\langle\tilde{f}_{-\vec{q}}\rangle$ and hence not trivially 0. To explicitly see this 
non-vanishing of  satellite asymmetry from disorder pinning, we discuss a simple model put forward in Ref.~\cite{ravy1992destructive}. Consider a single impurity at a random site $\vec{R}_0$ that interacts with the charge density such that the phase of the charge density is fixed to a value $\phi_0$ at site $\vec{R}_0$. The pinned CDW is given by $\rho_n=\rho_0 \sin\left[\vec{q}_c\cdot(\vec{R}_n-\vec{R}_{0})+\phi_0\right]$. The lattice modulations are in quadrature with the CDW and is given by $\vec{u}_n=\vec{u}_0 \cos\left[\vec{q}_c\cdot(\vec{R}_n-\vec{R}_{0})+\phi_0\right]$. Taking $f_I$ as the atomic form factor of the impurity and $f_0$ as that of the pure atom, the satellite asymmetry [Eq.~\eqref{Eq_asymmetry}] for this single impurity pinning gives,
\begin{eqnarray}
I(\vec{G}+\vec{q})-I(\vec{G}-\vec{q})=2N^{-1/2}f_{0}(f_I-f_0)(\vec{G}\cdot\vec{u}_0)\sin(\phi_0)
\end{eqnarray}
A maximum asymmetry is when $(f_I-f_0)\sin(\phi_0)=1$ which corresponds to the CDW having a maximum or minimum over the impurity depending on whether the interaction is attractive or repulsive. This picture describes strong pinning, where the CDW is pinned to a constant phase $\phi_0$ above each impurity. However, the pinning for a Bragg glass is weak,  where the phase is modulated by the collective interaction of impurities. A calculation of the asymmetry for Bragg glass was carried out in Refs.~\citenum{Rosso2003Phys.Rev.B} and \citenum{Rosso2004Phys.Rev.B}, and was shown to be an experimentally observable effect in principle.

\section{E: Momentum dependence of CDW Peak width}\label{SME}

In addition to the phase fluctuations that destroy long range CDW order, displacement of atoms from their ideal lattice sites (displacement  fluctuations) that destroy long range lattice order will also contribute to the broadening of the CDW peaks. Here we show that the width due to displacement fluctuations is momentum $(\vec{Q})$ dependent, in contrast to the $\vec{Q}$ independent broadening due to CDW phase fluctuations. Our model is similar to that of a paracrystal [chapter.~9 of Ref.\cite{Guinier1994}], with the modification of introducing a CDW with phase fluctuations on top of the lattice displacements. Using the same 3D lattice with $N$ sites as in SM-B, but with an additional lattice displacement $\vec{u}_n$ that can arise from thermal vibrations or disorder interaction,  the atoms are arranged at  $\vec{r}_n=\vec{R}_n+\vec{c}_n+\vec{u}_n$ where $\vec{R}_n$ are the  lattice sites and $\vec{c}_n$ are the CDW displacements. The scattering intensity at a momentum $\vec{Q}$ [Eq.~\eqref{Eq_IQ_phase}] is modified for the disordered lattice as, 
\begin{eqnarray}
I(\vec{Q})&=&\sum_{n,m}e^{i\vec{Q}\cdot(\vec{R}_n-\vec{R}_m)}\langle e^{i\vec{Q}\cdot(\vec{c}_n-\vec{c}_m)}\rangle_{_{\phi}}\langle e^{i\vec{Q}\cdot(\vec{u}_n-\vec{u}_m)}\rangle_{_{u}}\label{Eq_IQ_full}
\end{eqnarray}
where $\langle\cdots\rangle_{_{u}}$ denotes  ensemble average over lattice displacement configurations $\{u_n\}$, and we have assumed the lattice displacements are uncorrelated with the phase fluctuations. The CDW intensity around $\vec{Q}=\vec{G}+\vec{q}_c$ in Eq.~\eqref{Eq_Ic_sum} is now modified to
\begin{eqnarray}
I(\vec{Q}=\vec{G}+ \vec{q}_c+\delta\vec{q})&=&\dfrac{1}{4}(\vec{Q}\cdot\vec{c}_0)^2\sum_{n,m}e^{i\delta\vec{q}\cdot(\vec{R}_n-\vec{R}_m)}\langle e^{-i(\phi_n-\phi_m)}\rangle_{_{\phi}}\langle e^{-i\vec{Q}\cdot(\vec{u}_n-\vec{u}_m)}\rangle_{_{u}} \label{Eq_Ic_sum_un}
\end{eqnarray}
where the factor $\langle e^{-i\vec{Q}\cdot(\vec{u}_n-\vec{u}_m)}\rangle_{_{u}}$  under the Gaussian approximation gives 
\begin{eqnarray}
\langle e^{-i\vec{Q}\cdot(\vec{u}_n-\vec{u}_m)}\rangle_{_{u}}&=&   e^{-\frac{1}{2}\langle(\vec{Q}\cdot(\vec{u}_n-\vec{u}_m))^2\rangle_{_{u}}} + \mathcal{O}[\langle(\vec{Q}\cdot(\vec{u}_n-\vec{u}_m))^4\rangle_{_{u}}].\label{Eq_gauss_approx_un}
\end{eqnarray}
where $\langle(\vec{Q}\cdot(\vec{u}_n-\vec{u}_m))^2\rangle_{_{u}}$ quantify the mean squared fluctuations in relative lattice displacements. Defining a correlation function $C_{u}(\vec{r},\vec{Q})$ for the displacements relative to a reference point given by,
\begin{eqnarray}
C_{u}(\vec{r},\vec{Q})=e^{-\frac{1}{2}\langle\left(\vec{Q}\cdot(\vec{u}(\vec{r})-\vec{u}(0))\right)^2\rangle_{_{u}}},\label{Eq_Cu}
\end{eqnarray} 
where $\vec{u}(\vec{R}_n)\equiv \vec{u}_n$, the CDW peak intensity in Eq.~\eqref{Eq_Ic_integral} is modified to,
\begin{eqnarray}
I(\vec{Q}=\vec{G}+ \vec{q}_c+\delta\vec{q})\approx \dfrac{1}{4}(\vec{Q}\cdot\vec{c}_0)^2N\mathit{v}^{-1}\int (d^3\vec{r}) e^{i\delta\vec{q}\cdot\vec{r}}C_{\phi}(\vec{r})C_{u}(\vec{r},\vec{Q})\label{Eq_Ic_integral_un}
\end{eqnarray}

What sets the displacement fluctuations apart from CDW phase fluctuations is the $\vec{Q}$ dependence of $C_{u}(\vec{r},\vec{Q})$. It is this distinction that leads to the $\vec{Q}$ dependent broadening signature for the displacement fluctuations. To see this, consider an  exponentially decaying form for the displacement correlation given by  $C_{u}(\vec{r},\vec{Q})\sim e^{-|\vec{Q}|^2(\gamma_u r)}$. Here $\gamma_u$ with dimensions of length can be interpreted as the root mean square value of the relative displacement between neighboring atoms. When combined with the short range phase correlation $C_{\phi}(\vec{r})\sim e^{-r/\zeta_{\phi}}$,  Eq.~\eqref{Eq_Ic_integral_un}   gives an approximately Lorentzian peak profile at $\vec{Q}=\vec{G}\pm \vec{q}_c+\delta\vec{q}$ given by
\begin{eqnarray}
I(\vec{Q})\propto  \dfrac{1}{\left(1+\dfrac{|\delta\vec{q}|^2}{(\zeta_{\phi}^{-1}+|\vec{Q}|^2\gamma_u)^2}\right)^2}
\end{eqnarray}
whose full width at half maxima (FWHM) is given by
\begin{eqnarray}
\text{FWHM}&\propto& \zeta_{\phi}^{-1}+|\vec{Q}|^2\gamma_u
\end{eqnarray}
This shows the quadratic in momentum broadening due to displacement fluctuations. While the above form was obtained for a simple displacement correlation function that decay isotropically, a more general form for the broadening would be
\begin{eqnarray}
\text{FWHM}&\propto& \zeta_{\phi}^{-1}+\gamma_HQ_H^2+\gamma_KQ_K^2+\gamma_LQ_L^2
\end{eqnarray}
and we do not include terms like $Q_HQ_K$ etc. as they violate the reflection symmetry of the lattice. From a quadratic fit to the momentum dependence of the FWHM, the contribution from phase fluctuations: $\zeta_{\phi}^{-1}$ can be extracted as the intercept. 

The momentum dependence of the widths were studied in the 1980's but for quasi one dimensional ordered materials with short range order~\cite{Emery1978PRB,Heilmann1979PRB,Spal1980XraySS,Endres1982JPhys_Chem_Solids}. Such an analysis in 3D materials has so far remained a challenge due to the large number of peaks in the reciprocal space that need to be analyzed.  

\textbf{Numerical illustration of momentum dependent peak broadening}: To complement the above derivation, we numerically calculate the scattering intensity [Eq.~\eqref{Eq_IQ_full}] for a 1D lattice model with short range ordered CDW phase and lattice displacements. 

\begin{figure}
		\centering
		\includegraphics[width=0.95\linewidth]{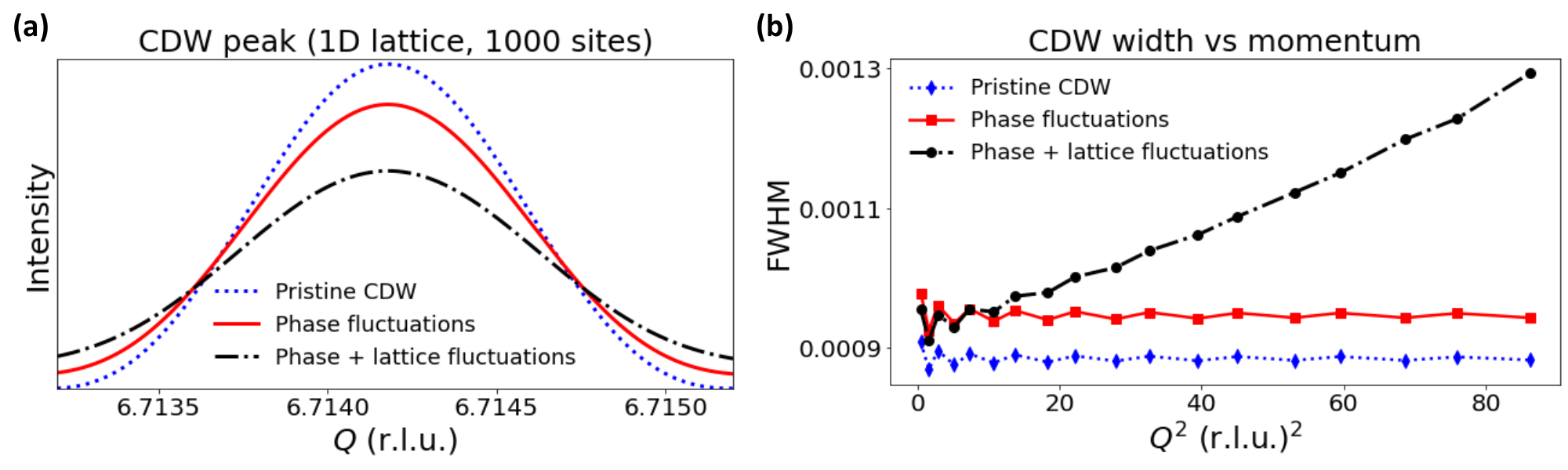}
		\caption{\textbf{(a):} Numerically calculated intensity of a CDW peak, and \textbf{(b):} the momentum $(Q)$ dependence of the full width at half maxima (FWHM) of the CDW peaks, in a 1D lattice with 1000 sites. The intensities are calculated for three disorder configurations: (1) pristine CDW with no disorder, (2) CDW with only phase fluctuations  and (3) CDW with both phase and lattice fluctuations. Both configurations (2) and (3) lead to a broadening of the peak [panel(a)]. However, the FWHM remains independent of $Q$ for the configuration with only phase fluctuations, while the FWHM for the configuration with both phase and lattice fluctuations show a $Q^2$ dependence. From the intercept of the $Q^2$ dependent FWHM, we can isolate the broadening contribution of the phase fluctuations.} 
		\label{fig:1D_simulation}
\end{figure}
On a lattice with 1000 sites, we set the CDW  modulation $q_c=2/7$ to mimic the CDW of RTe$_3$, and set the CDW amplitude = 0.01.  To generate a disordered phase configuration with short range correlation $C_{\phi}(|n-m|)\sim e^{-|n-m|/\zeta_{\phi}}$ between sites $n$ and $m$,  we start with the $n=0$ site where $\phi_{0}=0$ and the phases $\phi_n$ for each site $n>0$ are selected as $\phi_n=\phi_{n-1}+d\phi$ where $d\phi$ is drawn from a normal distribution with zero mean and standard deviation $\sigma_{\phi}$ (=0.025). This distribution generates phases whose mean square 
fluctuations are given by $\langle(\phi_n-\phi_m)^2\rangle_{\phi}=|n-m|\sigma_{\phi}^2$, and the phase correlation [Eq.~\eqref{Eq_Cr}] given by $C_{\phi}(|n-m|)=  e^{-|n-m|\sigma_{\phi}^2/2}$.  Similarly, to generate a short ranged lattice displacement configuration, the lattice displacements $u_n$ are generated as $u_n=u_{n-1}+du$ where $du$ is drawn from a normal distribution with zero mean and standard deviation $\sigma_{u}$ (=0.001), starting with $u_0=0$. This generates displacement configurations with mean squared fluctuation $\langle(u_n-u_m)^2\rangle_{u}=|n-m|\sigma_{u}^2$. We generate 400 realizations of phase and displacement configurations and calculate the intensity using Eq.~\eqref{Eq_IQ_full}. 

We show the calculated intensity profile of a CDW peak and the momentum ($Q$) dependence of the peak width in SM Fig.~\ref{fig:1D_simulation}. We see that while a short range ordered phase broadens the CDW peak whose width is independent of $Q$, a short range ordered lattice leads to broadening that is proportional to $Q^2$. In the presence of both short range ordered phase and lattice displacements, the $Q$ independent broadening due to phase only disorder can be extracted from the intercept of the $Q^2$ broadening.

\begin{figure}[t]
		\centering
		\includegraphics[width=0.95\linewidth]{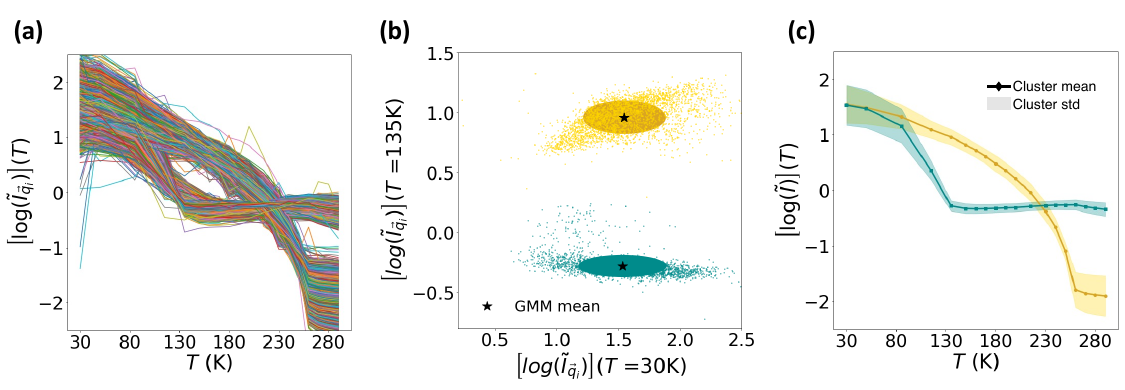}
		\caption{A simplified illustration of \xtec\ to cluster distinct intensity-temperature trajectories, $I(T)$, given the collection of series $\{I_{\vec{q}}(T_1), I_{\vec{q}}(T_2),\dots, I_{\vec{q}}(T_{d})\}$ ($d=19$ in this figure) at various momentum $\vec{q}$ in the reciprocal space. The raw trajectories rescaled as $\log[\tilde{I}_{\vec{q}}(T_i)]=\log[I_{\vec{q}}(T_i)]-\langle\log[I_{\vec{q}}(T_i)]\rangle_T$  [panel (a)] can be mapped to a simple Gaussian Mixture Model (GMM) clustering problem on a  $d$-dimensional space, whose 2D projection (along $T=30$K and $T=135$K) is shown in panel (b). The GMM identifies two distinct clusters and assigns them different colors. From the  cluster means (star symbol) and standard deviations (colored ellipsoids) of the GMM [panel (b)], we get the distinct trajectories of $\log[\tilde{I}(T)]$ and their standard deviation, with colors reflecting their cluster assignments [panel (c)]. 
} 
		\label{fig:XTEC_illustration}
\end{figure}

\begin{figure}[h]
    \includegraphics[width=0.8\linewidth]{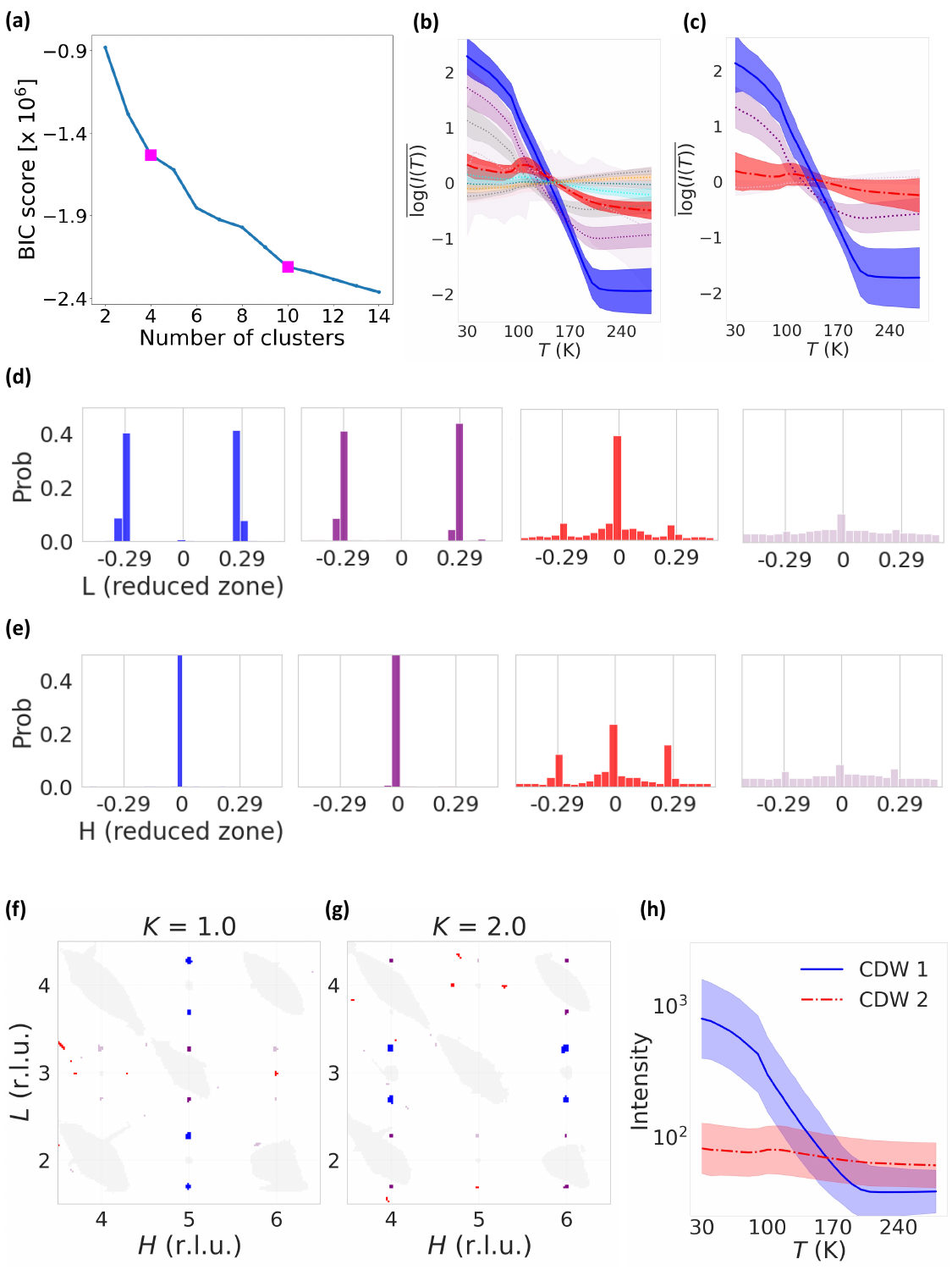}
    \centering
    \caption{Interpreting clusters for the x=2\% intercalation. \textbf{(a):} The BIC score for different numbers of clusters. Two heuristic estimates of 4 and 10 clusters are marked. \textbf{(b,c):} Cluster mean (lines) and variance as one standard deviation (shaded region) for GMM with 10 clusters in (b)  and 4 clusters in (c). \textbf{(d,e):} Probability distribution of the 4 clusters in (c) along $L$ (panel (d)) and $H$ (panel (e)) axis of the reduced Brillouin zone. \textbf{(f,g):} Distribution of the 4 clusters in the $H$-$L$ plane at $K=1$ (panel (f)) and $K=2$ (panel (g)). \textbf{(h):} Mean (lines) and one standard deviation (shading) of the intensity-temperature trajectories in the CDW-1 (blue) and CDW-2 (red) clusters. } 
    \label{Fig:XTEC_details}
 \end{figure}

\begin{figure}[h]
    \includegraphics[width=0.8\linewidth]{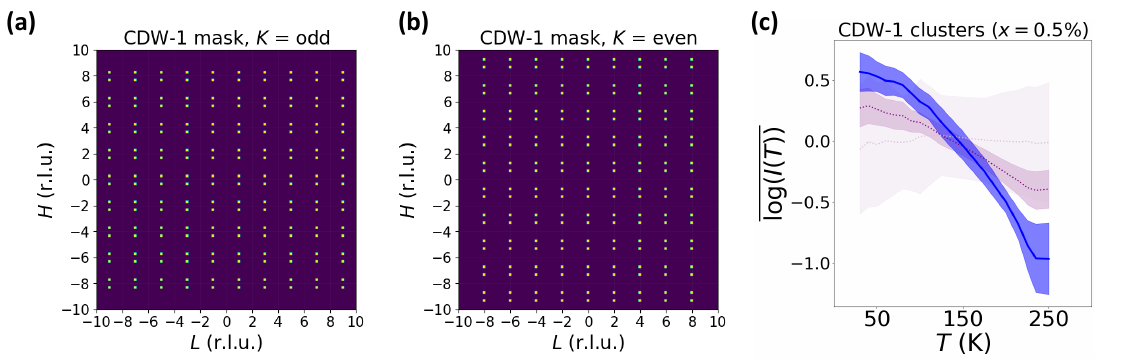}
    \centering
    \caption{CDW-1 masks for the \textbf{(a):} $K=$ odd planes and \textbf{(b):} $K=$ even planes. \textbf{(c):} The cluster mean (lines) and variance (shading) of the \xtec\ clustering with three clusters of CDW-1 trajectories filtered with the mask. This eliminates the weak and noisy peaks (purple and thistle colored) to isolate the CDW-1 peaks  with  well-defined trajectories (blue). }
    \label{Fig:CDW1_mask}
 \end{figure}

\section{F: X-ray Temperature Clustering: \xtec }\label{SMF}

The underlying principle of \xtec\ is to identify the distinct temperature trajectories through a Gaussian mixture model clustering~\cite{Venderley2022Proc.Natl.Acad.Sci.}. In SM-Fig.~\ref{fig:XTEC_illustration}, we show a simplified illustration of the GMM in action. A collection of raw intensity-temperature trajectories [SM-Fig.~\ref{fig:XTEC_illustration}(a)] given by $\{I_{\vec{q}}(T_1), I_{\vec{q}}(T_2),\dots, I_{\vec{q}}(T_{d})\}$  at various momenta $\vec{q}$ in reciprocal space can be represented as a distribution of points in a $d$ dimensional hyperspace, whose axis spans the intensities at each temperature. For visualization, a 2D cross-section of this hyperspace is shown in SM-Fig.~\ref{fig:XTEC_illustration}(b). The figure shows that the points are separated into two distinct groups (clusters). A Gaussian Mixture Model (GMM) clustering classifies these points into different clusters and assigns a mean and standard deviation for each cluster. The cluster mean reveals the distinct temperature trajectories in the data  [SM-Fig.~\ref{fig:XTEC_illustration}(c)], while the standard deviation shows that the clusters are well separated. In this example, a visual inspection of the raw intensities and a 2D projection can reveal the  distinct clusters. However, the real data is messier [See Fig.~1(e)] and requires a GMM clustering on the entire hyperspace to identify the distinct trajectories.

\subsection{XTEC analysis of intercalated samples}
In the Fig.~2 (a-c) of the main text, we benchmarked the XTEC clustering on the pristine sample. In this section  [and SM Fig.~\ref{Fig:XTEC_details}], we provide further details on    
the X-TEC analysis, using $x=2\%$ intercalation as a representative example. We list the details in the following steps,
\begin{enumerate}
    \item All the intensity slices in the $(H,L)$ plane with integer $K$ values are loaded. The first preprocessing step is the automated thresholding that removes the low-intensity background noise, as explained in Ref~\cite{Venderley2022Proc.Natl.Acad.Sci.}. 
    \item Next, we implement XTEC with label smoothing (XTEC-s) through peak averaging (Ref~\cite{Venderley2022Proc.Natl.Acad.Sci.}). For this,each set of connected pixels (that passed the thresholding) in the reciprocal space is identified as a single peak. The intensity of each peak is given by its peak average value. This step removes the resolution-limited pixel-to-pixel fluctuations in the intensity. After this step, $\sim100,000$ non-Bragg peaks are identified, which include CDWs, detector artifacts, and background scattering. 
    \item The peak averaged intensities are rescaled as $\overline{\log[I_{\vec{q}}(T_i)]}=\log[I_{\vec{q}}(T_i)]-\langle\log[I_{\vec{q}}(T_i)]\rangle_T$. This step ensures that the  clustering reveals the distinct intensity-temperature trajectories rather than the absolute magnitude of intensities.

    \item The next step is to identify the optimal number of clusters. A Bayesian information criterion (BIC)~\cite{bishop2006pattern,scikit-learn} can provide a heuristic estimate of the optimal number. For the 2\% intercalation shown in SM Fig~\ref{Fig:XTEC_details}(a), the elbow method roughly points to $4$ or $10$ clusters. To move forward, a physicist's  intervention is required to identify and interpret each cluster. \end{enumerate}
SM Fig.~\ref{Fig:XTEC_details} (b) and (c) show the clustered trajectories given  by the cluster mean and one standard deviation of each cluster. From the trajectories, we see that 4 clusters are sufficient to reveal all the unique trajectories. To understand the physical significance of each cluster, we analyze the location of each cluster in the momentum space. For the 2\% intercalation, as shown in SM Fig.~\ref{Fig:XTEC_details} (d,e), we see that the blue and purple cluster is predominantly made of pixels located at $(H,L) \equiv (0,0.29)$ corresponding to CDW-1, while the red clusters are located at $(H,L) \equiv (0.29,0)$ corresponding to CDW-2. The last cluster (thistle colored) has no characteristic location in the momentum space and corresponds to the background intensity. The 
 reciprocal space with the pixels assigned their cluster colors [SM Fig.~\ref{Fig:XTEC_details} (f,g)] shows that CDW-1 and CDW-2 are arranged in the same 3D pattern as that of the pristine sample [Main Fig.~2 (b,c)]. The blue and red clusters identify with the primary CDW-1 and CDW-2 peaks, respectively, while the purple cluster captures the higher-order peaks of CDW-1. Having identified which pixel correspond to the CDW peaks, we can look at the average intensity of each cluster (the peak height of CDWs) [SM Fig.~\ref{Fig:XTEC_details} (h)]. Compared to the pristine sample [Main Fig.~2 (a)], a 2\% intercalation strongly suppresses the CDW-2 with no sharp onset behavior.

\subsection{Filtering the CDW-1 peaks}
In the previous section, we showed a brute-force XTEC analysis on the full data.  That analysis identified the CDW peaks and their 3D structure in the reciprocal space. We can now more precisely target the CDW-1 peaks by filtering out only the pixels of the primary CDW-1 peaks with a mask. We apply the mask shown in SM Fig~\ref{Fig:CDW1_mask}(a,b) on the raw intensities before feeding to XTEC. We select the mask with sufficiently large windows to capture broad peaks such that the results are robust to the size of the mask. The filtered intensities are then fed into the XTEC pipeline, as described in the previous section. A clustering on these  filtered intensities [SM Fig~\ref{Fig:CDW1_mask} (c)] eliminates the background as well as the noisy and weak CDW-1 peaks (purple and thistle clusters). This procedure gives us a collection of  $\sim 3,000$ intense and well-defined CDW-1 peaks (blue cluster) that can be robustly analyzed to extract their peak height and spread. The results in main Fig.~2(d-f) and Fig.~3 follow this analysis.

\begin{figure}[t]
    \includegraphics[width=0.8\linewidth]{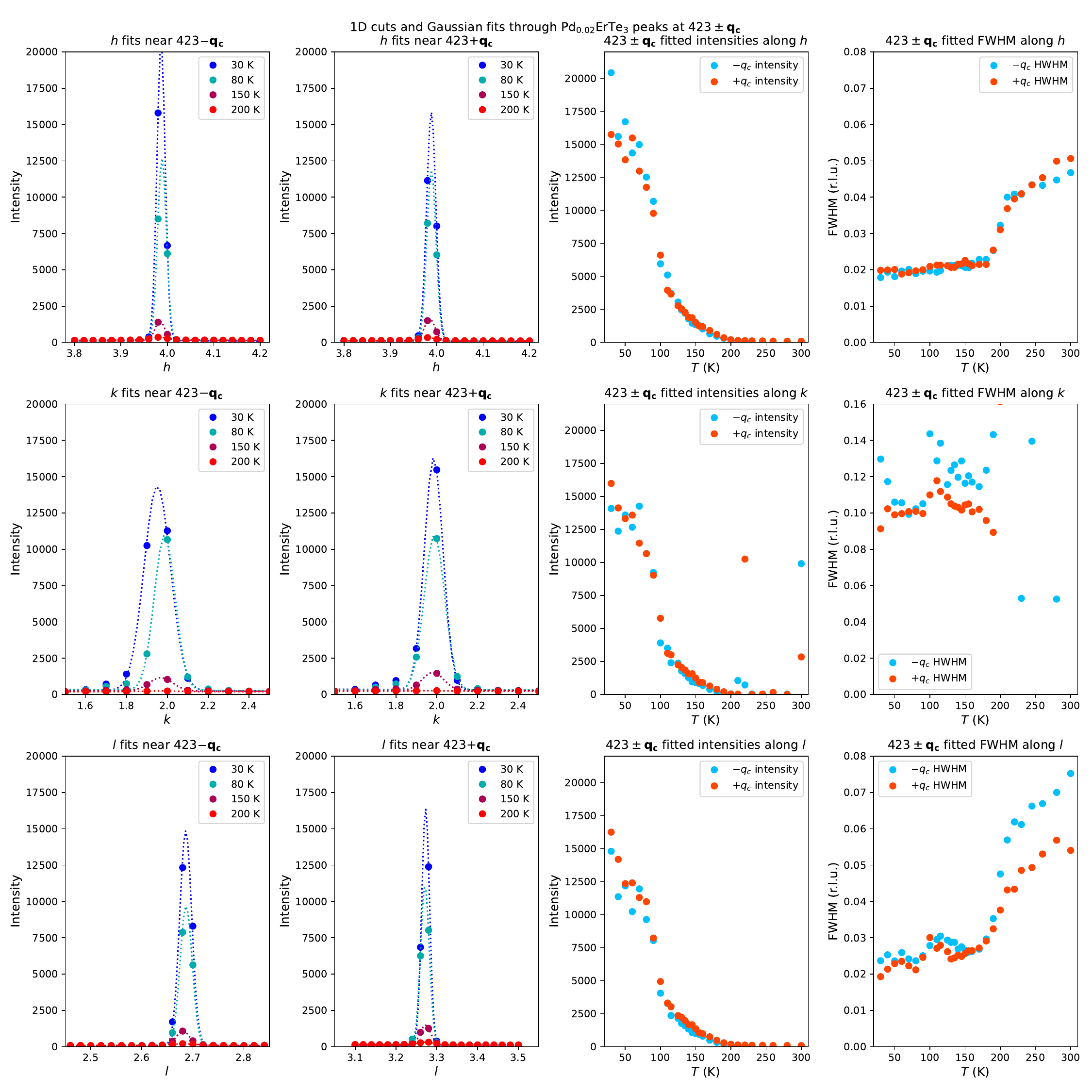}
    \centering
    \caption{A set of three one-dimensional Gaussian functions can be used to fit superstructure peaks and extract intensity and width parameters to characterize peaks. This approach is applied here to the $432\pm \mathbf{q}_c, \mathbf{q}_c \approx \frac{2}{7} \mathbf{c}^*$ superstructure peaks in Pd$_{0.02}$ErTe$_3$. For all rows, the left two figures are cuts taken through superstructure peaks, with circles indicating data and dotted lines representing fitted Gaussians; the right two figures are fitted parameters at different temperatures. \textbf{Top row}: Line cuts taken along $h$. \textbf{Middle row}: Line cuts taken along $k$. Note that the peak become so broad along $k$ above $T\approx 210$ K that the fitting function fails.  \textbf{Bottom row}: Line cuts taken along $l$.  It is notable that the full-width half maximum (FWHM) of the fit approaches the bin width at low temperatures, indicating that the peak is resolution-limited.}\label{SM_fig:1D_Gauss_fits}
\end{figure}

\section{G: Conventional peak width analysis}\label{SMG}

The conventional approach of extracting the peak width would be to take one-dimensional line cuts through CDW peaks in the binned data and extract intensity and width parameters from fitting.  Fitting domains must be chosen arbitrarily in relation to diffuse scattering and spurious crystallographic imperfections, making this approach difficult to apply to the entire dataset. Moreover, the necessity of determining the goodness of fit makes this approach impractical to scale. Applying this approach to an ad-hoc choice of peaks at $hkl = 4\ 2\ 3\pm q_c$ in SM Fig.~\ref{SM_fig:1D_Gauss_fits} for the $2\%$ intercalation shows that the  spurious signals make it difficult to find a uniform way to fit even a small number of peaks in these data, and even the best-fitted parameters will have low precision. It is also clear from SM Fig.~\ref{SM_fig:1D_Gauss_fits} that it is impossible to extract any power law tails from fitting these peaks. Thus, relying on power law tails as a signature of Bragg glass is not feasible.

The challenge in estimating the subtle peak height asymmetry or the profile asymmetry  of the CDW satellite peaks  in intercalated samples is also clear from SM Fig.~\ref{SM_fig:1D_Gauss_fits}. The peak intensities from the Gaussian fit on two pairs of satellite CDW-1 peaks do not reveal a stark asymmetry. Comparing the magnitude of the peak intensity ($\sim$20,000 counts) and the strength of the asymmetry seen in the diffuse scattering ($\sim$10 counts, see main Fig 4), the scale of asymmetry is too small to be detected from the fluctuations of peak height intensity. The peak widths of the two satellite peaks are also nearly identical, and hence we do not detect a clear signature of profile asymmetry. 

\begin{figure}
    \includegraphics[width=0.8\linewidth]{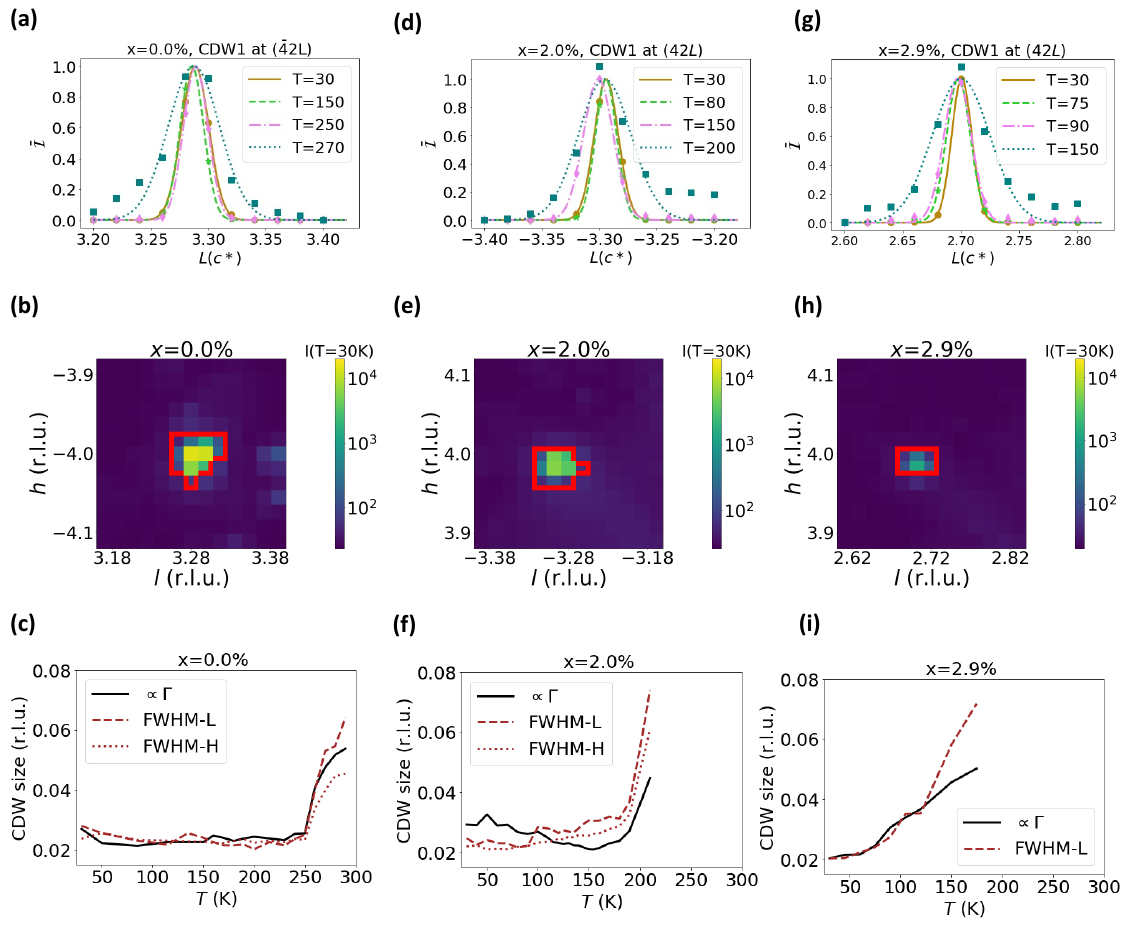}
    \centering
    \caption{Benchmarking the peak spread. \textbf{(a):} Line cut along a  CDW-1 peak at various temperatures $T$ for the pristine sample ($x=0\%$). The intensities (symbols) are averaged along $H=\bar{4}\pm 0.02$ r.l.u and $K=2$, and normalized with its maximum value at the peak. The minimum intensity in the line cut is subtracted to remove any background offset. The lines are Gaussian fits.  \textbf{(b):} The intensity of the CDW-1 peak in the $K=2$ plane [same peak as in (a)] at $T=30$K, with the \xtec\ determined peak boundaries (red contour) for the $x=0\%$.   \textbf{(c):} The peak spread  ($\Gamma$) [Eq.~(1) of main text] for the CDW peak in (a-b), along with the FWHM from line cuts along $H$ (FWHM-H) and $L$ (FWHM-L), at various $T$ for $x=0\%$. \textbf{(d,e,f):} Same as panels (a-c) respectively but for $x=2\%$ intercalated sample.  \textbf{(g,h,i):} Same as panels (a-c)  but for $x=2.9\%$ intercalated sample.}\label{SM_fig:Gamma_vs_linecut}
\end{figure}

\begin{figure}
\includegraphics[width=0.8\linewidth]{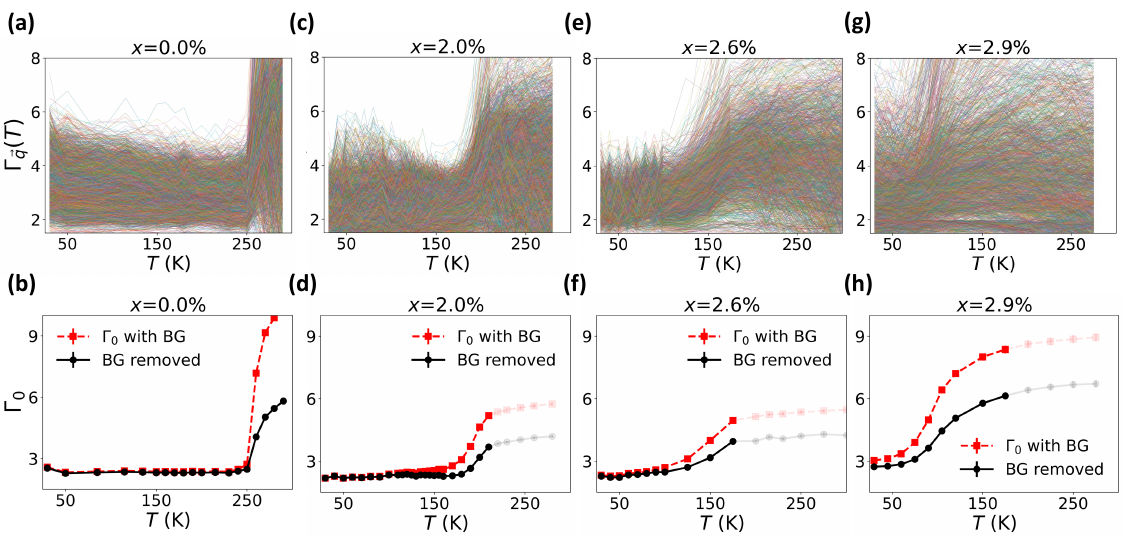}
    \centering
    \caption{Extracting the peak spread. \textbf{(a):} The $T$ trajectory of the peak spread ($\Gamma_{\vec{q}}$) of 4877 CDW-1 peaks in the $x=0\%$ data (thin colored lines). \textbf{(b):} Comparing $\Gamma_0$ with the background intensity offset removed (BG removed), and $\Gamma_0$ keeping the background offset ($\Gamma_0$ with BG). Lines are guides to the eyes. \textbf{(c,d):} Same as (a,b) respectively but for $x=2\%$ sample with 3688 peaks. The peak spread saturates above a cutoff temperature.  \textbf{(e,f):} $x=2.6\%$ intercalated sample with 2251 peaks. \textbf{(g,h):} $x=2.9\%$ with 2713 peaks. In panels (d), (f) and (h), the peak spread saturates above a cutoff temperature, when the peak intensity drops near the background value.}\label{SM_fig:Gamma_vs_background}
\end{figure}

\section{H: Extracting Peak spread analysis with \xtec}\label{SMH}

In this section, we first provide details of the steps to extract the  peak spread $\Gamma_{\vec{q}}(T)$ [Eq.~(1) of main text] from the XRD data and benchmark them with line cuts on selected CDW-1 peaks [SM Fig.~\ref{SM_fig:Gamma_vs_linecut}]. We then show the underlying quadratic momentum dependence of $\Gamma_{\vec{q}}$, and the extraction of the $\vec{q}$ independent term $\Gamma_0$ that quantifies the broadening purely due to CDW phase fluctuations [SM Fig.~\ref{SM_fig:Gamma_vs_momentum}].

\subsection{1. Benchmarking peak spread }
The conventional approach to extract a FWHM is shown for a CDW-1 peak at the three levels of intercalation in SM Fig.~\ref{SM_fig:Gamma_vs_linecut} (a), (d) and (g). Our high throughput measure $\Gamma_{\vec{q}}$ [Eq.~(1) of main text] directly provides a measure for the spread of the peak (in units of the number of pixels). This is achieved by using \xtec\ to identify the connected pixels whose intensity trajectories belong to the CDW order parameter cluster. This is shown in SM Fig.~\ref{SM_fig:Gamma_vs_linecut} (b), (e) and (h), where the red boundary determined by \xtec\ marks the extent of the CDW peaks.  We quantify the spread of this CDW peak (centered at momentum $\vec{q}$)  with $\Gamma_{\vec{q}}$ which is the ratio of the total intensity inside the peak boundary to the maximum intensity of the peak. We restrict to the in-plane peak spread with intensities  at integer $K$ values of the out of plane ($b^*$) axis, to avoid the lower resolution along $b^*$ axis [0.1 (r.l.u.) compared to 0.02 (r.l.u.) for the in-plane] from limiting the overall resolution of the spread.  The estimated $\Gamma_{\vec{q}}$ is compared with the FWHMs of the line cuts in SM Fig.~\ref{SM_fig:Gamma_vs_linecut} (c), (g) and (i). We see that the $\Gamma_{\vec{q}}$ faithfully captures the features of the FWHMs, in particular, the rapid onset of broadening above a transition temperature. 

\begin{figure}[t!]
		\centering
		\includegraphics[width=0.95\linewidth]{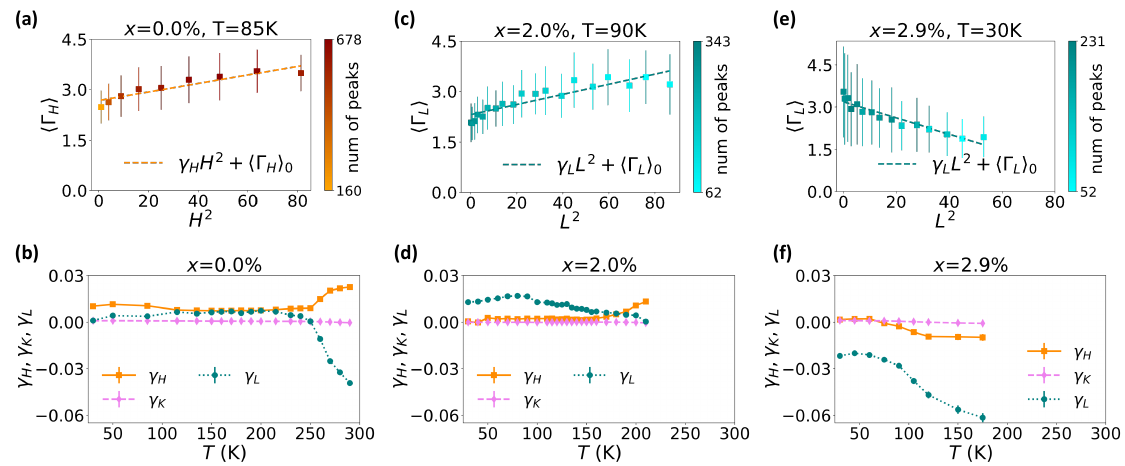}
		\caption{\textbf{(a)}:  The $H^2$ dependence of the peak spread in the $x=0\%$ sample at $T$=85K. The $\langle\Gamma_H\rangle$ (symbols) is the spread obtained by averaging  $\Gamma_{\vec{q}}$  over $K$ and $L$ that share the same $|H|$. The error bars indicate standard deviation of $\Gamma_{\vec{q}}$ at $|H|$. The markers are color-coded, and the color bar indicates the number of peaks determining the statistics of each marker.  The fit: $\gamma_H H^2+\Gamma_0$ agrees well with $\langle\Gamma_H\rangle$ within the error markers.  \textbf{(b)}: The  momentum coefficients $\gamma_H$, $\gamma_K$, $\gamma_L$ from the 3D quadratic fit [Eq.~(2) of main text] to $\{\Gamma_{\vec{q}}\}$ at various temperatures for the $x=0\%$. The lines are guides to the eye.  \textbf{(c)}: Same as (a) but for the  $L^2$ dependence of the spread, $\langle\Gamma_L\rangle$ (symbols)  (by averaging  $\Gamma_{\vec{q}}$  over $H$ and $K$ at $|L|$) for the $x=2\%$ intercalated sample at $T$=90K. The fit $\gamma_L L^2+\langle\Gamma_L\rangle_0$ also agrees well within the standard deviation of $\Gamma_{\vec{q}}$ at $L$. \textbf{(d)}: Same as (b), but for the $x=2\%$ intercalated sample. \textbf{(e-f)} Same as (c-d) respectively, but for $x=2.9\%$ intercalated sample.} 
		\label{SM_fig:Gamma_vs_momentum}
\end{figure}

However, both the FWHMs and the $\Gamma_{\vec{q}}$ show an erratic temperature trajectory, reflecting the errors in the width estimation from the small resolution peaks (the peaks are roughly spread over 2-3 pixels). Collecting all $\Gamma_{\vec{q}}$ with $\vec{q}$ spanning $\sim 3000$ peaks, we find a wide variation in the range of values for the spread, [see SM Fig.~\ref{SM_fig:Gamma_vs_background} (a), (c), (e) and (g)]. Buried under these seemingly erratic trajectories is the systematic $\vec{q}$ dependence from lattice distortions [see SM Fig.~\ref{SM_fig:Gamma_vs_momentum}] and the unique $\vec{q}$ independent spread $\Gamma_0$ of the disordered CDW [see Fig.~3 of main text].

An important step in the estimation of $\Gamma_{\vec{q}}$ is the removal of the background intensity offset from the CDW peak intensities. This background contribution is estimated as the average of the intensities outside the CDW boundary in a 10x10 pixel neighborhood of the peak [the blue region outside the red boundary in SM Fig.~\ref{SM_fig:Gamma_vs_background} (b), (d), (f) and (h)], and this offset contribution is subtracted from the total and maximum intensity of the CDW peak before estimating $\Gamma_{q}$.  In SM Fig.~\ref{SM_fig:Gamma_vs_background} (b), (d), (f) and (h), we show the effect of not removing the background offset on the $\Gamma_{0}$. Keeping the background intensity results in an overestimation of the spread, especially at higher temperatures where the peak height is smaller. This is because the peak spread measure misinterprets the extra background intensity outside the true peak as a genuine broadening of the peak. 

 In SM Fig.~\ref{SM_fig:Gamma_vs_background} (d), (f) and (h), the peak spread saturates at high temperatures above a cut-off temperature. Above this  temperature, the peak spreads beyond the boundaries of the peak determined by \xtec\ [SM Fig.~\ref{SM_fig:Gamma_vs_linecut} (b), (e), (h)]. This is also the point where the peak intensity drops low enough to a value near the background intensity (see Fig.~2(d) of main text).

\begin{figure}[t]
    \includegraphics[width=0.7\linewidth]{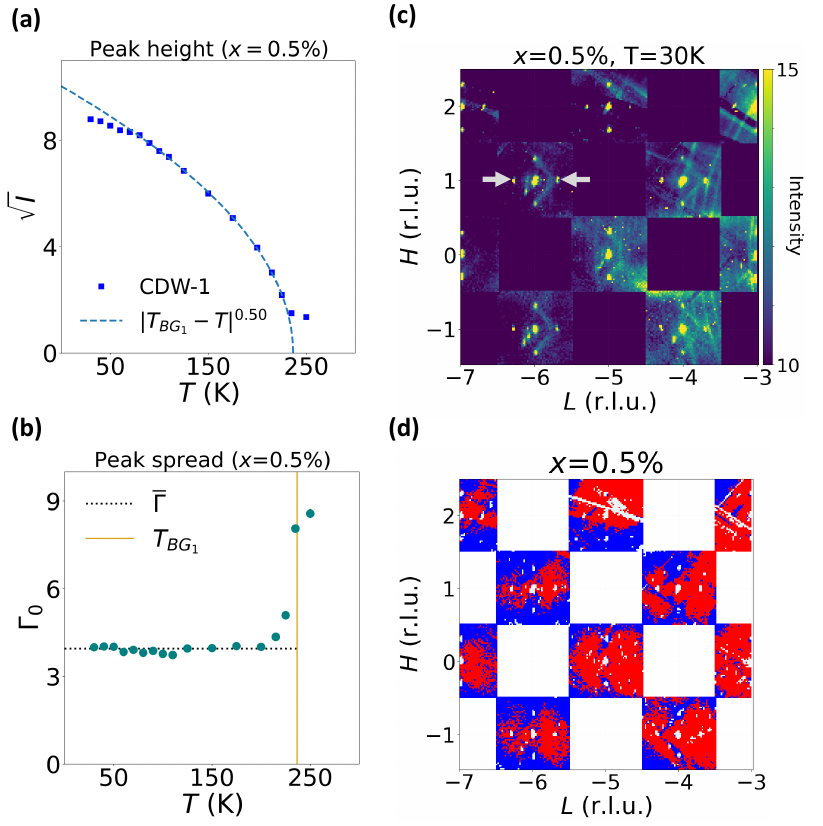}
    \centering
    \caption{x=0.5\% sample. {\bf{(a):}} The CDW-1 peak averaged intensity (peak height). The $\tilde{I}$ is obtained from the average of all the intensities in the CDW-1 cluster (2323 peaks), from which we subtract the background intensity contribution. $\sqrt{\tilde{I}(T)}$  fits well to a power law $\propto(T_{BG_1}-T)^{\beta}$ giving  $T_{BG_1}\sim235$K and $\beta=0.5$ matching the BCS order parameter exponent. {\bf{(b):}} The $\vec{q}$ independent broadening of CDW-1 peak spread, $\Gamma_0(T)$, extracted from 2323 peaks by fitting their $\Gamma_{\vec{q}}$ to a quadratic function of $\vec{q}$ [Eq.~2 of main text]. The $T_{BG_1}=235$K extracted from panel (a) is also shown. {\bf{(c):}}  The intensity at $T=30$K, in the $H$-$L$ plane with $K=1.0$, shows a diffuse scattering that is asymmetrically distributed between the two satellite peaks, in the form of half diamonds. (see arrows for reference) {\bf{(d):}} Two cluster \xtec\ results color coded as red and blue, from the temperature trajectories  of the diffuse scattering intensities. The pixels are colored red (blue) if their intensity trajectory  belongs to the red (blue) cluster. The intensities of the CDW peaks and $H$+$L$ = odd Bragg peaks  (white pixels, identified from a prior \xtec \ analysis)  are excluded from this two-cluster \xtec, along with the $H$+$L$ = even Bragg peaks removed by a square mask (square white regions) }
    \label{Fig:0.5data}
 \end{figure}

\subsection{2. Momentum dependence of peak spread }

In SM Fig.~\ref{SM_fig:Gamma_vs_momentum}, we show that the spread $\Gamma_{\vec{q}}$ in the XRD data has a systematic broadening with quadratic dependence in $\vec{q}$ as predicted in SM-E. To simplify the visualization of the 3D quadratic fit in Eq.~(2) of main text, we project the momentum dependence of $\Gamma_{\vec{q}}$ to one direction by  averaging over the other directions. We show the $H^2$ dependence of $\Gamma_{\vec{q}}$ in SM Fig.~\ref{SM_fig:Gamma_vs_momentum} (a), and the $L^2$ dependence in (c) and (e), for the three levels of intercalation.   The $\Gamma_{\vec{q}}$ fits well with the quadratic function.

The full 3D fit of $\Gamma_{\vec{q}}$ using Eq.~(2) of main text extracts the quadratic coefficients $\gamma_H$, $\gamma_K$, $\gamma_L$ as well as  the momentum-independent intercept $\Gamma_0$. While $\Gamma_0$ is reported in the main text [Fig.~3 (c-f)], in SM Fig.~\ref{SM_fig:Gamma_vs_momentum} (b), (d) and (f) we report their respective $\gamma_H$, $\gamma_K$, and $\gamma_L$ values.

\section{I: Bragg glass in $x=0.5$\% intercalation.}
We show the analysis of 0.5\% sample separate from the main figures, as this sample showed a much larger mosaic spread compared to all other samples. In SM Fig.~\ref{Fig:0.5data}, we show the filtered CDW-1 peaks of the 0.5\% intercalation. 

Unlike the other intercalated samples (2\%, 2.6\%, and 2.9\%), the 0.5\% is much similar to the pristine sample, and shows the BCS scaling order parameter Fig.~\ref{Fig:0.5data}(a), with a critical temperature of $235K$. The peak spread is shown in SM Fig.~\ref{Fig:0.5data}(b). Despite the similarity with the pristine sample, $x=0.5$\% is distinguished by the presence of asymmetry in the diffuse scattering [SM Fig.~\ref{Fig:0.5data}(c,d)]. The presence of the distinct half-diamond asymmetry, similar to that of other intercalated samples, shows that the 0.5\% intercalation is disordered and different from the pristine sample. Thus, long range order is forbidden in this sample, and the transition temperature at $235K$ corresponds to a Bragg glass transition.

\begin{figure}[t]
    \includegraphics[width=0.9\linewidth]{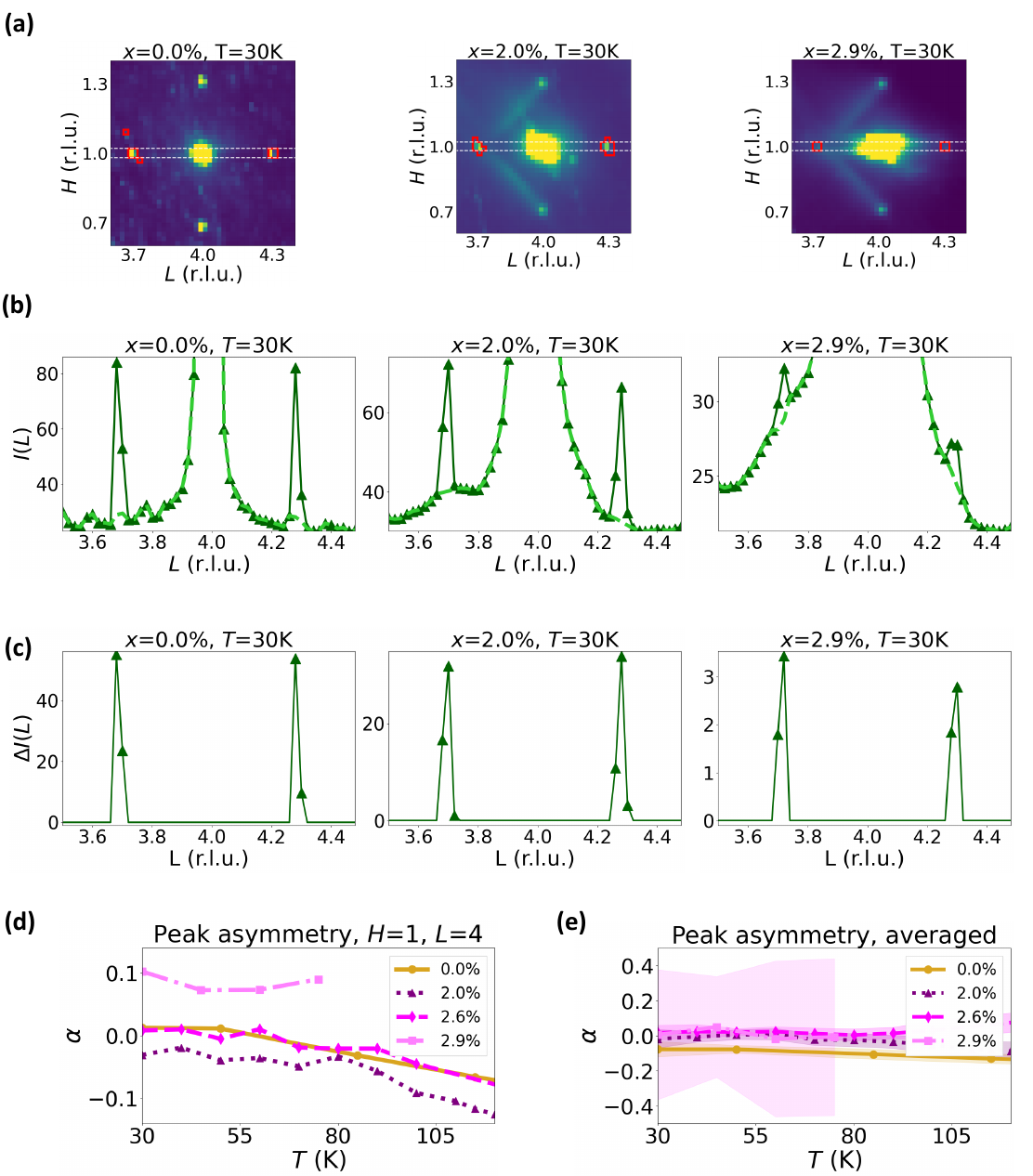}
    \centering
    \caption{CDW satellite peak asymmetry. {\bf{(a):}} The in-plane XRD  near the $(H=1, L=4)$ Bragg peak at $T=30$K, with $K$ (out-of-plane axis) averaged over all values between $-20$ and $20$ (r.l.u). The pixels identified by X-TEC as CDW-1 peaks are enclosed within the red boundary. The horizontal dashed line marks the region considered for the line cuts in panel (b).  {\bf{(b):}} Line-cut intensities across the CDW-1 peaks along $L$ with intensity averaged over $H\in [0.98,1.02]$ (r.l.u) and $K\in[-20,20]$ (r.l.u). The solid line shows the raw intensity, and the dashed line shows the interpolated intensities after removing the CDW pixels. {\bf{(c):}} The CDW-1 peak intensities after subtracting the interpolated (dashed line) from the raw intensities (solid line) of the panel (b). {\bf{(d):}}  Asymmetry in the satellite peaks  surrounding $(H=1, L=4)$, quantified by the ratio $\alpha=(I_L-I_R)/(I_L+I_R)$, where $I_L$ and $I_R$ are the heights of the left and right CDW-1 peak intensities. {\bf{(e):}} Mean (lines) and standard deviation (shading) of $\alpha$ from all pairs of CDW satellite peaks in the $-1.5\leq H\leq 2.5$ and $3\leq L\leq 8.5$.}
    \label{Fig:SM_Peak_asymmetry}
 \end{figure}
\section{J: CDW satellite peak asymmetry.}
In Fig 4 of the main text, we discussed the asymmetry in the diffuse scattering surrounding the CDW peaks. The diffuse scattering asymmetry is present only in intercalated samples, and clearly distinguishes them as disordered. In this section and in SM Fig~\ref{Fig:SM_Peak_asymmetry}, we investigate the asymmetry in the CDW peaks after removing the diffuse background. As will be apparent from the following discussion, this analysis is prone to errors from the pixelation of peaks and the interpolation of background diffuse scattering. Hence these results should be interpreted with a grain of salt.

We implement a punch-and-fill method to isolate the CDW peak heights from the diffuse background. As shown in SM Fig.~\ref{Fig:SM_Peak_asymmetry} (a), we identify the pixels of the CDW peaks  (enclosed within the red boundary) with \xtec. These CDW pixels are punched out and filled with a linear spline interpolation of their neighboring intensities. A line cut through the CDW-1 satellite peaks shows the CDW peaks (solid lines) and the interpolated background (dashed line) in SM Fig.~\ref{Fig:SM_Peak_asymmetry} (b). Subtracting the two isolates the CDW peaks from the diffuse scattering.

The background removed in-plane intensities of CDW-1 satellite peaks around the Bragg peak at $(H,L) = (1,4)$ and $K$ averaged along all values in $[-20,20]$ is shown in SM Fig.~\ref{Fig:SM_Peak_asymmetry} (c). Unlike the stark asymmetry apparent in the background, the asymmetry is not obvious in the CDW peaks and is subtle, if any, even for the strongest intercalation. Moreover, with only two data points to span the peaks, it is impossible to fit the peak profile.

Using the peak maxima as the estimate for the peak height, the asymmetry for the peaks around $(H,L) = (1,4)$ is quantified in  SM Fig.~\ref{Fig:SM_Peak_asymmetry} (d),  as $\alpha=(I_L-I_R)/(I_L+I_R)$ where $I_L$ and $I_R$ is the peak maxima of the left and right CDW peak.  A noticeable asymmetry is seen for the 2.9\% intercalation. However, the very small height of peaks of 2.9\% (an order of magnitude smaller than the pristine and 2\% intercalation) makes them more sensitive to errors in subtracting the background.  

For a more accurate and comprehensive estimate of the asymmetry ratio, in Fig.~\ref{Fig:SM_Peak_asymmetry} (e), we measure the peak heights in the 2D $(H,L)$ plane rather than along line-cuts. The panel shows the mean and standard deviation of asymmetry from all pairs of in-plane CDW satellite peak intensities in the $-1.5\leq H\leq 2.5$ and $3\leq L\leq 8.5$ (same region shown in Fig 4 (a-b) of main text). In this region, the asymmetry ratio of only those pairs of peaks that are present in all four samples are chosen. The figure suggests that intercalated samples have a slightly higher value of asymmetry than the pristine sample. However, the overall small values of the ratio and their large variance, as well as the susceptibility of the analysis to pixelation and background subtraction  errors, prevent us from establishing conclusive evidence of asymmetry in the peak heights, even at the highest intercalation (x=2.9\%) which is a clearly disordered sample.  However, the diffuse asymmetry that surrounds the CDW peaks already makes the total peak height asymmetric [Fig.~4(c-e) of main text]. This is a clear sign that intercalation introduces disorder pinning to the modulations at and around the CDW.

\begin{figure}[t]
    \includegraphics[width=0.9\linewidth]{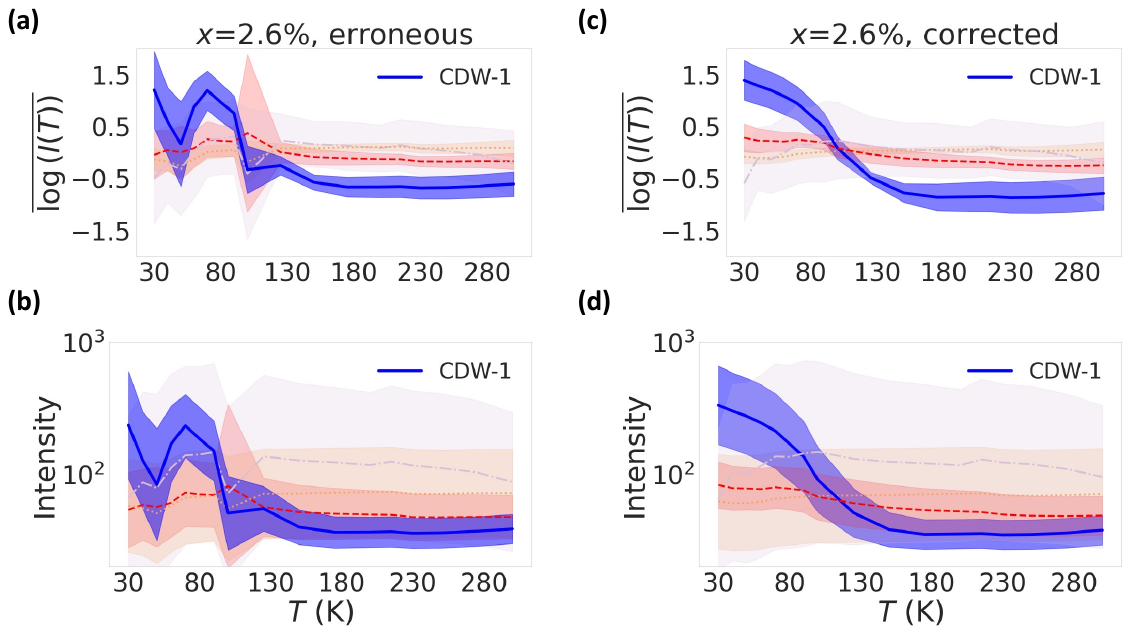}
    \centering
    \caption{ {\bf{(a,b):}} The \xtec\ analysis on 2.6\% intercalation, with a slight misorientation error during the image transformation to reciprocal space bins. The trajectory of all clusters including the blue CDW-1 cluster show sudden discontinuities. Panel (a) shows the cluster mean (lines) and variance (shading) of the GMM with rescaled intensities. Panel (b) shows the average and one standard deviation of the unscaled intensities in each cluster. {\bf{(c,d):}} Same as panels (a) and (b) but for the correctly oriented transformation of the same 2.6\% intercalation. }
    \label{Fig:SM_wrong_binning}
 \end{figure}

\section{K: Caveats with the improper orientation of XRD transformation.}
In this section, we discuss the importance of ensuring the correct orientation at each temperature during the transformation of images to the reciprocal space bins. In SM Fig:\ref{Fig:SM_wrong_binning}, we show an instance of improper orientation on the 2.6\% sample. In panels (a-b), we see the blue \xtec\ cluster that identify with the CDW-1 have an anomalous hump, and an erratic trajectory. Sharp jumps in temperature are seen concurrently for all the clusters, including the CDW-1 cluster. 

A more careful transformation of the same 2.6\% data, where the reciprocal space images at different temperatures are ensured to have the same correct orientation, eliminates the erratic trajectories and sharp jumps. In SM Fig:\ref{Fig:SM_wrong_binning} (c,d), the \xtec\ analysis shows smooth trajectories for the CDW-1 (blue cluster) as well as the other clusters of the diffuse background. 

The \xtec\ analysis thus  makes it easy for the scientist to  visually inspect orientation errors during the transformation of images to reciprocal space. A misorientation will show up as sharp discontinuities occurring concurrently in the majority of the clusters. These discontinuities should not be confused with a first-order phase transition, as  in the latter; only the cluster that captures the order parameter shows the discontinuity, while the remaining clusters have smooth trajectories.

\end{document}